\documentclass[preprint,12pt]{elsarticle}

\usepackage{amssymb}
\usepackage{amsmath}
\usepackage{pifont}
\usepackage{url}
\usepackage[utf8]{inputenc}
\usepackage[T1]{fontenc}
\DeclareUnicodeCharacter{00F6}{\"{o}}
\usepackage{microtype}
\usepackage[colorlinks=true,allcolors=blue]{hyperref}

\usepackage{nomencl}
\makenomenclature
\usepackage{etoolbox}
\renewcommand\nomgroup[1]{%
  \item[\bfseries
  \ifstrequal{#1}{O}{Symbols}{%
  \ifstrequal{#1}{A}{Acronym}{%
  \ifstrequal{#1}{S}{Subscripts and superscript}{%
  \ifstrequal{#1}{G}{Greeks Letters}{}}}}%
]}

 \usepackage{geometry}

\journal{Energy Conversion and Management}

\begin{document}

\begin{frontmatter}



\title{Parasitic Hydrogen Bubble Evolution in Vanadium Redox Flow Batteries: A Lattice Boltzmann Study} 


\author[1,2]{K. Duan} 

\author[1,2]{T.H. Vu}
\author[1]{T. Kadyk\corref{cor1}}
\ead{t.kadyk@fz-juelich.de}
\author[3]{Q. Xie}
\author[3,4]{J. Harting}
\author[1,2]{M. Eikerling}

\cortext[cor1]{Corresponding author}

\affiliation[1]{organization={Institute of Energy Technologies, IET-3: Theory and Computation of Energy Materials, Forschungszentrum Jülich GmbH},
            addressline={Wilhelm-Johnen-Straße}, 
            city={Jülich},
            postcode={52425},
            country={Germany}}

\affiliation[2]{
            organization={Theory and Computation of Energy Materials, Faculty of Georesources and Materials Engineering, RWTH Aachen University},
            city={Aachen},
            postcode={52062},
            country={Germany}
            }
\affiliation[3]{
            organization={Helmholtz Institute Erlangen-Nürnberg for Renewable Energy (IET-2), Forschungszentrum Jülich GmbH},
            addressline={Cauerstr.~1}, 
            city={Erlangen},
            postcode={91058}, 
            country={Germany}
            }
\affiliation[4]{
            organization={Department of Chemical and Biological Engineering and Department of Physics, Friedrich-Alexander-Universität Erlangen-Nürnberg},
            addressline={Cauerstr.~1}, 
            city={Erlangen},
            postcode={91058}, 
            country={Germany}
            }
            
\begin{abstract}

Vanadium redox flow batteries (VRFBs) are promising for large-scale energy storage due to their long cycle life and flexible scalability. Their performance, however, can be compromised during charging by side reactions, most notably the hydrogen evolution reaction (HER) at the negative electrode, which generates gas bubbles that block electrolyte pathways and reduce electrochemically active area. Here, we present the first three-dimensional simulations of HER-driven bubble dynamics in $\mu$-CT–derived carbon felt electrodes using a color-gradient lattice Boltzmann method, which enables stable interface tracking and reduced spurious currents. Simulations show that higher HER rates intensify bubble growth and cause premature detachment, insufficient flow rates lead to persistent accumulation, and increased electrode compression hinders bubble removal despite improving conductivity. These findings highlight the need to balance gas suppression, flow conditions, and electrode design to effectively manage bubble formation and transport. The mechanistic insights gained here not only advance the understanding of HER-driven multiphase transport in VRFBs but also provide general guidance for optimizing porous electrode systems that rely on carbon felts.

\end{abstract}

\begin{graphicalabstract}
\includegraphics[width=1\linewidth]{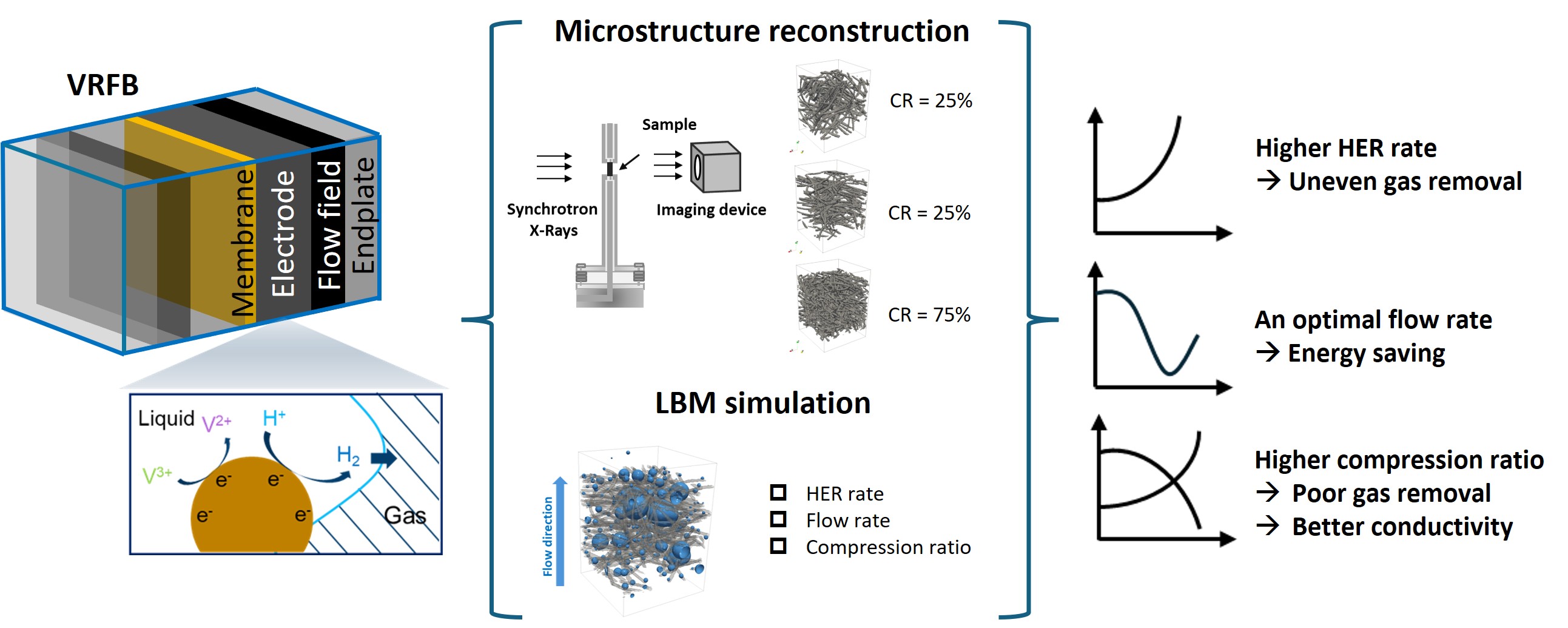}
\end{graphicalabstract}

\begin{highlights}
\item Uneven gas removal occurs with increased gas production in the electrode
\item Optimal flow rate minimizes bubbles and reduces external pumping energy
\item Higher compression hinders gas removal but boosts electrical conductivity
\end{highlights}


\end{frontmatter}


\nomenclature[O]{\(f_i\)}{distribution function for density in direction $i$}
\nomenclature[O]{\(F_i\)}{discrete force in direction $i$}
\nomenclature[O]{\(F\)}{body force}
\nomenclature[O]{\(e_i\)}{discrete velocity in direction $i$}
\nomenclature[O]{\(x\)}{lattice site position}
\nomenclature[O]{\(t\)}{lattice time}
\nomenclature[O]{\(c_s\)}{speed of sound}
\nomenclature[O]{\(u\)}{velocity}
\nomenclature[O]{\(P\)}{pressure}
\nomenclature[O]{\(G\)}{color gradient}
\nomenclature[O]{\(A\)}{parameter that controls the interfacial tension}
\nomenclature[O]{\(B_i\)}{weights to ensure mass conservation}
\nomenclature[O]{\(k_r\)}{reaction rate}
\nomenclature[O]{\(C_l\)}{length scale conversion factor}
\nomenclature[O]{\(C_t\)}{time scale conversion factor}
\nomenclature[O]{\(C_m\)}{mass scale conversion factor}
\nomenclature[O]{\(w_i\)}{weighting factor in direction $i$}

\nomenclature[A]{\(VRFBs\)}{vanadium redox flow batteries}
\nomenclature[A]{\(HER\)}{hydrogen evolution reaction}
\nomenclature[A]{\(SHE\)}{standard hydrogen potential}
\nomenclature[A]{\(LBM\)}{lattice Boltzmann method}
\nomenclature[A]{\(CT\)}{computerized tomography}
\nomenclature[A]{\(KARA\)}{Karlsruhe Research Accelerator}
\nomenclature[A]{\(CR\)}{compression ratio}
\nomenclature[A]{\(BGK\)}{Bhatnagar-Gross-Krook}
\nomenclature[A]{\(l.u.\)}{lattice unit}

\nomenclature[G]{\(\Omega_i\)}{collision operator in direction $i$}
\nomenclature[G]{\(\rho\)}{density}
\nomenclature[G]{\(\rho_0\)}{reference density}
\nomenclature[G]{\(\phi\)}{color function}
\nomenclature[G]{\(\alpha\)}{fluid component}
\nomenclature[G]{\(\sigma\)}{interfacial tension}
\nomenclature[G]{\(\tau\)}{relaxation time}
\nomenclature[G]{\(\beta\)}{parameter that controls the interface thickness}
\nomenclature[G]{\(\theta_i\)}{the angle between color gradient and lattice discrete velocity}
\nomenclature[G]{\(\bar\nu\)}{mixed fluid viscosity}
\nomenclature[G]{\(\nu_r\)}{red fluid kinematic viscosity}
\nomenclature[G]{\(\nu_b\)}{blue fluid kinematic viscosity}

\nomenclature[S]{\(0\)}{initial values}
\nomenclature[S]{\(i\)}{discrete velocity direction}
\nomenclature[S]{\(r\)}{red fluid}
\nomenclature[S]{\(b\)}{blue fluid}
\nomenclature[S]{\(BGK\)}{Bhatnagar-Gross-Krook}
\nomenclature[S]{\(pert\)}{perturbation}
\nomenclature[S]{\(rec\)}{recoloring}
\nomenclature[S]{\(eq\)}{equilibrium}
\nomenclature[S]{\(boundary\)}{lattice sites at the boundary}
\nomenclature[S]{\(LB\)}{lattice Boltzmann units}
\nomenclature[S]{\(Phys\)}{physical units}

\printnomenclature

\section{\label{sec:level1}Introduction}
\begin{figure}[t]
\centering
\includegraphics[width=1\linewidth]{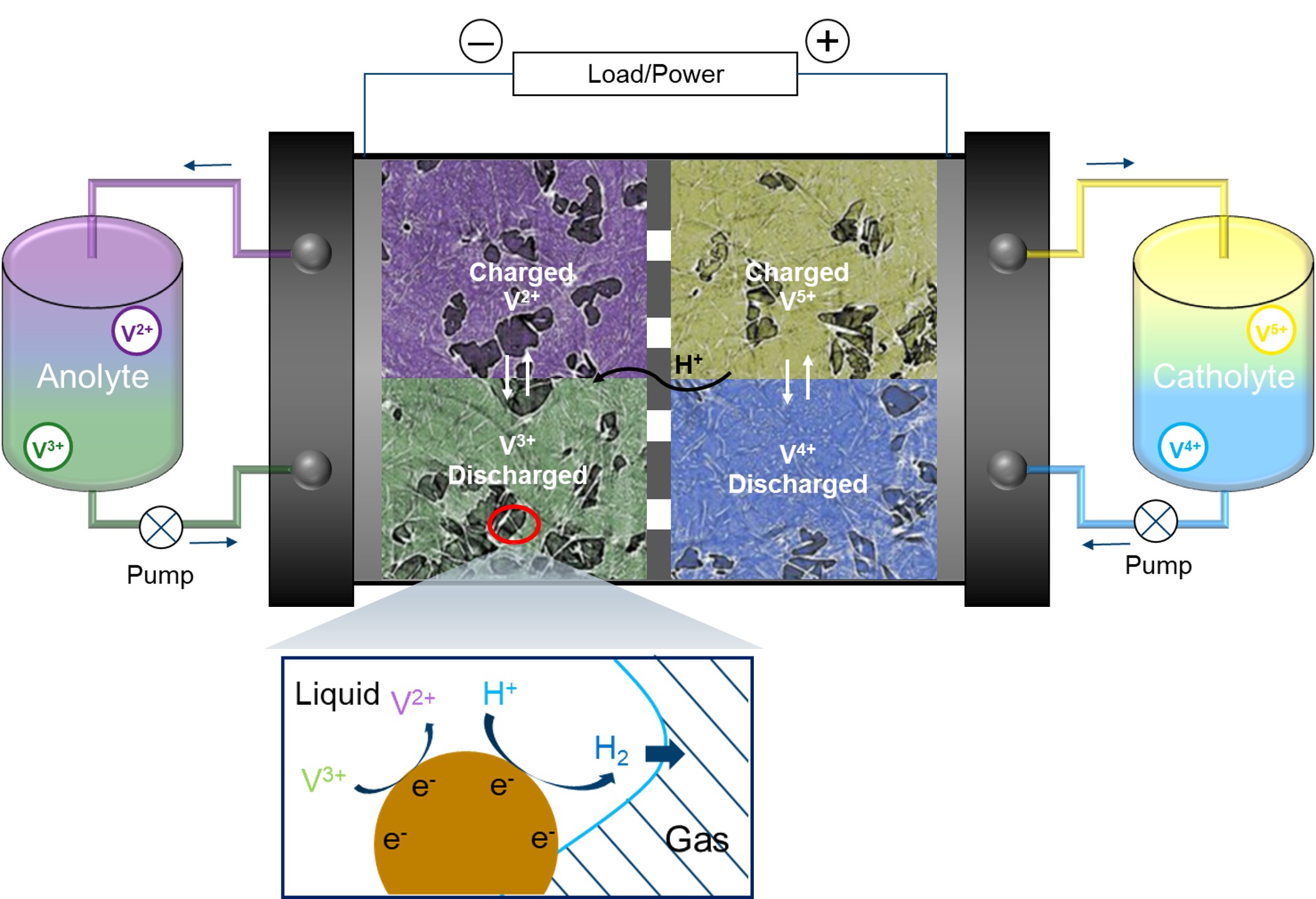} 
\caption{An illustration of the working principle for a vanadium redox flow battery.}
\label{fig:working_principle}
\end{figure}

Large-scale energy storage is essential for integrating intermittent renewable power sources, such as wind and solar, into modern grids~\cite{dunn2011electrical, skyllas2011progress,gur2018review}. Among available technologies, redox flow batteries (RFBs) offer unique scalability in both power and capacity, as well as long cycle life~\cite{Guney2017,Lourenssen2019,zhang2019progress}. Vanadium redox flow batteries (VRFBs), in particular, employ a single element in multiple oxidation states on both electrodes, thereby avoiding cross-contamination and ensuring long-term chemical stability~\cite{Ryan2019,Kear2012}. In the negative half-cell, the active species are V$^{2+}$/V$^{3+}$, while the positive half-cell contains V$^{4+}$/V$^{5+}$ couples~\cite{minke2018materials, shi2019recent}. These reactions enable highly reversible cycling (Fig.\ref{fig:working_principle}). Despite these advantages, VRFBs face persistent challenges that hinder widespread deployment, including low energy density, high material costs (vanadium salts, ion-exchange membranes, carbon electrodes), and complex maintenance procedures for long-term operation\cite{li2022techno,baritto2022development, rodby2020assessing}.

While VRFBs are designed as single-phase electrochemical systems, gas bubbles are frequently observed in porous electrodes under practical operating conditions. They may originate from residual air entrained during electrolyte filling~\cite{Koble2021,Bevilacqua2019} and, more importantly, from parasitic side reactions within the operating voltage window, such as hydrogen evolution, oxygen evolution, and carbon corrosion~\cite{Lourenssen2019,Huang2022}. Among these, the hydrogen evolution reaction (HER) on the negative electrode is particularly problematic due to its relatively low overpotential. When the electrode potential becomes sufficiently negative during charging, protons in the acidic electrolyte are reduced to hydrogen gas~\cite{Schweiss2016}:
\begin{align}
 \mathrm{2\,H^+ + 2\,e^- \rightarrow H_2 \uparrow} &\;\;\;\;  E^\circ \leq 0 \; \mathrm{V\;vs.\;SHE}  
\end{align}
Since the HER competes with the V$^{2+}$/V$^{3+}$ couple ($E^\circ = -0.255$ $\mathrm{V\;vs.\;SHE}$)~\cite{wu2018electrocatalysis, wang2018reduction}, it consumes charge, generates gas bubbles that hinder electrolyte transport, reduce electrochemically active area, and increase mass transport resistance~\cite{Ye2017,Ye2020,Zhao2019}. Over time, such parasitic processes degrade Coulombic efficiency, energy efficiency, and electrode microstructure.

Efforts to mitigate bubble-related performance losses have followed two main directions. The first suppresses gas generation by modifying electrode surfaces~\cite{AlNajjar2024,Wen2023,Schneider2019}, adjusting electrolyte composition~\cite{Choi2017,Tian2023,SkyllasKazacos2016,Wu2014}, or optimizing operating conditions~\cite{Fetyan2019,Zhang2015,Wei2017}. Najjar et al.~\cite{AlNajjar2024} used tungsten oxide to modify carbon cloth electrodes on the negative half-cell. The results showed that the V$^{2+}$/V$^{3+}$ reaction kinetics was significantly improved compared to untreated carbon cloth, and the HER was inhibited. The second promotes rapid bubble removal through hydrodynamic design or surface engineering, including tailored flow channels~\cite{Eifert2020, chai2024double}, wettability modification~\cite{Ye2020,Koble2023}, and active bubble pumping~\cite{Ye2024, ma2024evaluation}. These approaches improve bubble detachment and limit blockage of electrolyte pathways. Chai et al.~\cite{chai2024double} designed a new double-spiral flow channel to significantly reduce the pressure drop in VRFB system. Ma et al.~\cite{ma2024evaluation} found that hydrogen bubble generation significantly increases pressure drop and pump power consumption. Ye et al.~\cite{Ye2024} reported that increasing pumping pressure beyond a threshold alleviated flow choking by detaching bubbles from pores, whereas K\"oble et al.~\cite{Koble2023} found that iron-doped carbon–nitrogen materials improved wettability and electrolyte saturation.

However, the dynamics of bubble evolution inside realistic electrode microstructures remain poorly understood. Existing operando imaging studies, such as synchrotron X-ray tomography~\cite{Eifert2020,Koble2024, duan2025investigating,colliard2025advancing}, capture macroscopic gas distributions but cannot resolve local fiber–bubble interactions. Furthermore, the high cost and scarcity of such experimental facilities hinder systematic investigation. Numerical simulation offers a complementary pathway, with the lattice Boltzmann method (LBM) being particularly well suited for modeling multiphase transport in complex porous structures~\cite{Chen2017,Zhang2018}. Previous LBM studies, such as those by Chen et al.~\cite{Chen2017} using the Shan–Chen pseudo-potential model~\cite{shan1993lattice}, have examined the effects of porosity, fiber diameter, wettability, and saturation on gas migration by imposing an initial random two-phase distribution. Zhang et al.~\cite{Zhang2018} investigated the role of wetted area on battery performance in a similar framework. However, such methods overlook the complete progression of HER-driven bubble dynamics, from nucleation to detachment, and are prone to higher spurious currents compared to more advanced multiphase models.

To the best of our knowledge, this work is the first study to simulate HER‑induced bubble dynamics in $\mu$‑CT–derived carbon felt geometries across multiple compression ratios (CRs) using a three‑dimensional color‑gradient LBM framework. This modeling framework provides improved numerical stability, accurate control of surface tension, and reduced spurious currents compared to the Shan–Chen model~\cite{Leclaire2017b,scheel2023viscous}, enabling direct simulation of bubble nucleation and growth under reactive flow conditions. By systematically varying reaction rate, electrolyte velocity, and electrode compression, we elucidate how microstructural characteristics and operating parameters jointly influence bubble evolution and transport. Section~\ref{section2} describes the electrode geometries and modeling methodology, Section~\ref{section3} presents the parametric analysis of bubble behavior, and Section~\ref{section4} concludes with implications for the design and operation of VRFB electrodes.

\section{Methodology}\label{section2}

\subsection{Images and reconstructions}

The experimental procedure utilizes synchrotron-based X-ray micro-CT imaging to examine a GFA 6EA carbon felt electrode, part of the SIGRACELL\textsuperscript{\textregistered} battery electrode series manufactured by SGL Carbon in Germany. This non-destructive method allows for high-resolution, three-dimensional visualization of the electrode's internal structure. Imaging takes place at the Karlsruhe Research Accelerator (KARA) in Germany. Detailed information about the imaging approach can be found in Ref.~\citenum{Bevilacqua2019}.

\begin{figure*}[htbp]
\centering
\includegraphics[width=1\linewidth]{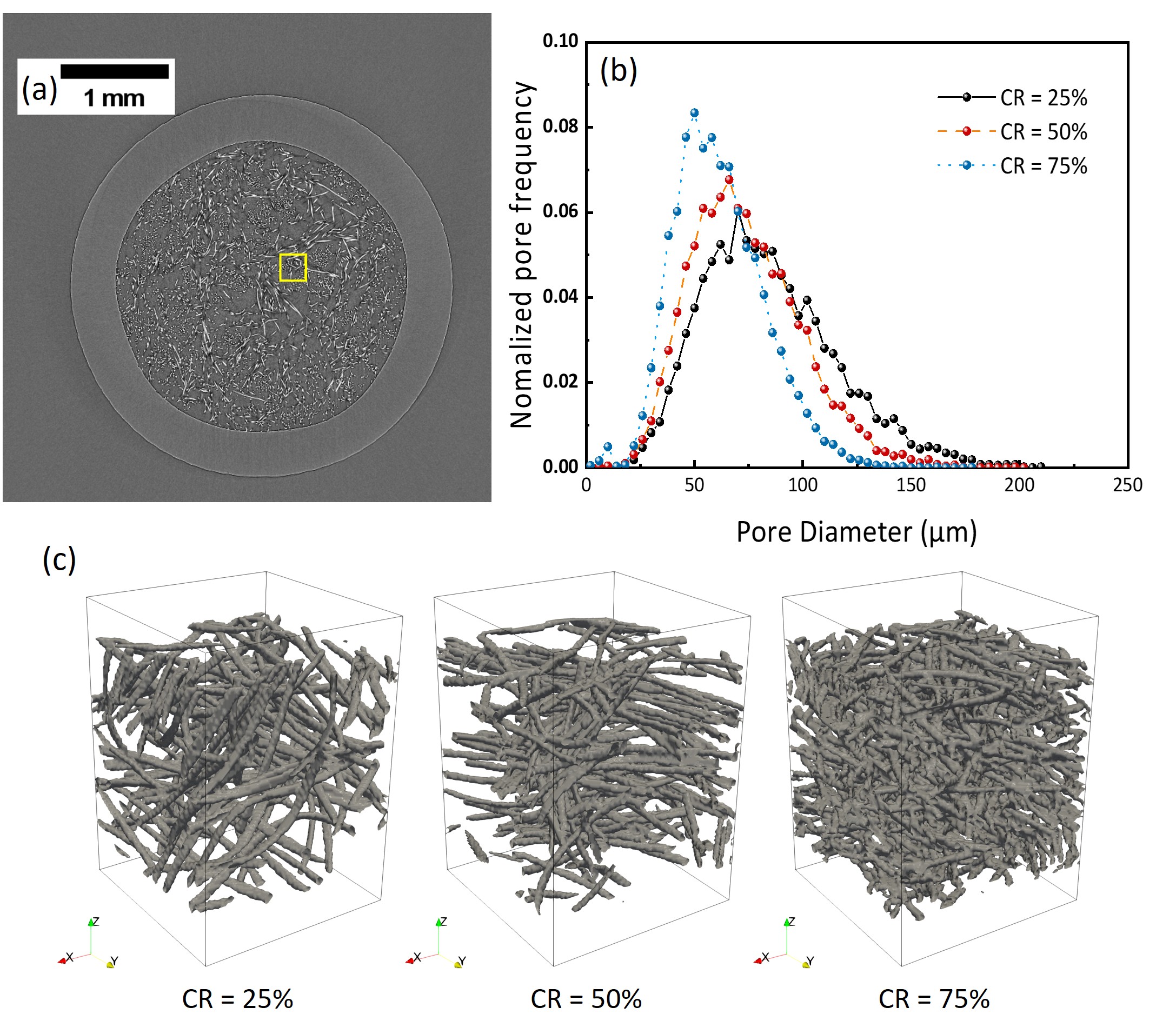} 
\caption{ (a) A synchrotron tomography image (the yellow box marks the selected region); (b) pore size distribution with different compression ratios; (c) 3D visualizations of different microstructures corresponding to different compression ratios.}
\label{fig:reconstruction}
\end{figure*}

The carbon felt electrode is captured with a resolution of $2015 \times 2015 \times 2015$ voxels, with each voxel representing $2.44^{3}\,\mu\text{m}^{3}$, as shown in Fig.~\ref{fig:reconstruction}(a). We select a cube of size of $100^{3}$ voxels from the area near the center of the image set to construct the microstructures used in the present study (as shown in the yellow box in Fig.~\ref{fig:reconstruction}(a)). The grey ring around the porous structure is the tube that holds the electrodes in the experiment~\cite{Bevilacqua2019}. The carbon fibers within the felt have an average diameter of about 10~$\mu$m. Fig.~\ref{fig:reconstruction}(b) shows the pore size distribution, under three different CRs. CR is defined as the ratio of the compressed thickness to the original thickness of the electrode, and it ranges from 25\% to 75\%. Fig.~\ref{fig:reconstruction}(c) is a three-dimensional visualization of the electrode structure used in the present study, where the grey phase represents the carbon fibers.

It should be noted that while the present analysis focuses on a specific commercial felt and three representative compression ratios, the general morphology of carbon felts is highly heterogeneous. Variations in porosity, wettability, and fiber orientation among different electrode types may influence bubble behavior and quantitative outcomes. Nevertheless, the reconstructed structures employed here capture the key features of fibrous porous media, allowing mechanistic insights into bubble evolution that are expected to extend to a broad class of carbon-felt–based electrodes.

\subsection{Lattice Boltzmann method for immiscible two-phase flow}

A three-dimensional 19 velocity (D3Q19) color-gradient lattice Boltzmann model is adapted to investigate the immiscible two-phase flow. The approach is based on the work by Leclaire et al. \cite{Leclaire2017b,scheel2023viscous}. The evolution equation for the single particle distribution function $f_i(\mathbf{x},t)$ is given by
\begin{equation}
f_i(\mathbf{x}+\mathbf{e}_i\,\delta t,\,t+\delta t)= f_i(\mathbf{x},t) + \Omega_i + F_i, 
\end{equation}

\noindent
where $\mathbf{x}$ denoted the position of the lattice site and $t$ is the time, $\mathbf{e}_i$ is the discrete velocity in the $i$-th direction, $\Omega_i$ is the collision operator, and $F_i$ represents the forcing term. Here, the time increment $\delta t$ and lattice spacing $\delta x$ are taken as unity. In this model, two sets of distribution functions, $f_i^r$ (red) and $f_i^b$ (blue), are introduced to represent the two different fluids. The total distribution function is defined as
\begin{equation}
f_i = f_i^r + f_i^b. 
\end{equation}
\noindent
The collision operator $\Omega_i$ consists of three parts,
\begin{equation}
\Omega_i = \Omega_i^{\mathrm{BGK}} + \Omega_i^{\mathrm{pert}} + \Omega_i^{\mathrm{rec}}, 
\end{equation}
\noindent
where $\Omega_i^{\mathrm{BGK}}$ is the single-phase collision operator, $\Omega_i^{\mathrm{pert}}$ is the perturbation operator which generates an interfacial tension, and $\Omega_i^{\mathrm{rec}}$ is the recoloring operator used to enforce phase segregation and to maintain an interface between components.

We employ the simple and popular Bhatnagar-Gross-Krook (BGK) collision operator, i.e.,
\begin{equation}
\Omega_i^{\mathrm{BGK}} = -\frac{1}{\tau}\left(f_i - f_i^{\mathrm{eq}}\right), 
\end{equation}
\noindent
where  $\tau = 3\,\bar\nu + 0.5$  is the effective relaxation time. The mixed fluid viscosity $\bar\nu$ is defined via a harmonic density-weighted average of the red fluid kinematic viscosity $\nu_r$ and the blue fluid kinematic viscosity $\nu_b$. The equilibrium distribution function $f_i^{\mathrm{eq}}$ is obtained by expanding the Maxwell–Boltzmann distribution in a Taylor series of the local fluid velocity $\mathbf{u}$ up to second order,
\begin{equation}
f_i^{\mathrm{eq}} = w_i\,\rho \left[ 1 + \frac{\mathbf{e}_i\cdot\mathbf{u}}{c_s^2} + \frac{(\mathbf{e}_i\cdot\mathbf{u})^2}{2c_s^4} - \frac{\mathbf{u}\cdot\mathbf{u}}{2c_s^2} \right], 
\end{equation}
\noindent
where $c_s=1/\sqrt{3}$ is the speed of sound, $w_i$ is a weighting factor depending on the lattice direction ($w_0=1/3$, $w_{1-6}=1/18$, $w_{7-18}=1/36$), and $\rho$ is the total density, with $\rho = \rho^{r}+\rho^{b}$, and $\rho^{r}$, $\rho^{b}$ being the densities of red and blue fluids, respectively. The density of each fluid is given by the zeroth moment of its distribution functions,
\begin{equation}
\rho^{\alpha} = \sum_i f_i^\alpha, \quad \alpha = r,\, b, 
\end{equation}
\noindent
and their reference densities, $\rho_0^r$ and $\rho_0^b$, are both set to 1 in lattice units (l.u.) in the present study. The total density is defined as the first moment of the color-blind distribution functions,
\begin{equation}
\rho\,\mathbf{u} = \sum_i f_i\,\mathbf{e}_i \;+\; \frac{\delta_{t}}{2}\,\mathbf{F}, 
\end{equation}
\noindent
and the pressure $\mathbf{P}$ is calculated as follows:
\begin{equation}
\mathbf{P}=c_s^{2}\,\rho.  
\end{equation}
In the present study, a pressure gradient is introduced to drive the flow, which is modeled as a body force $\mathbf{F}$ based on Guo's forcing scheme \cite{guo2002discrete}. 
Phase separation is implemented using the color gradient method, which introduces an interaction between different fluid components resulting in the separation of phases in three steps \cite{Leclaire2017b, scheel2023viscous,gunstensen1991lattice}.  In the first step, the direction of the steepest increase in the density of the respective fluid component (the color gradient $\mathbf{G}$) is calculated as
\begin{equation}
\mathbf{G} = \nabla \phi = \frac{3}{\delta t} \sum_i \,w_i\, \mathbf{e}_i\, \phi(t,\mathbf{x}+\mathbf{e}_i\,\delta t),
\end{equation}
\noindent
where $\phi$ is a so-called color function (order parameter) used to identify the interface. It is defined by
\begin{equation}
\phi = \frac{\rho^{r} - \rho^{b}}{\rho^{r} + \rho^{b}}. 
\end{equation}
In the subsequent perturbation step, populations aligned with the gradient of the density field of fluid component $\alpha$ are increased, while those perpendicular to the gradient are diminished, resulting in the emergence of a surface tension,
\begin{equation}
\Omega_i^{\mathrm{pert}} = A\,\lvert \mathbf{G} \rvert
\biggr[
\,w_i\frac{(\mathbf{e}_i\cdot \mathbf{G})^2}{\lvert \mathbf{G} \rvert^2}
- B_i
\biggr], 
\end{equation}
\noindent
where the parameter $A = \frac{9\,\sigma}{4\tau}$ (space- and time-dependent) controls the interfacial tension $\sigma$ at the fluid interface, and the weights $B_i$ are chosen to ensure mass conservation ($B_0=2/9$, $B_{1-6}=1/54$, $B_{7-18}=1/27$).
In the final step, known as the recoloring step, the two phases are separated by redistributing the two fluid populations in opposite directions, that is
\begin{align}
f_i^r &= \frac{\rho^r}{\rho} \,f_i + \beta \,\frac{\rho^r \,\rho^b}{\rho^2} \,\cos(\theta_i) \,f_i^{\mathrm{eq}}(\rho, 0),  \\
f_i^b &= \frac{\rho^b}{\rho} \,f_i - \beta \,\frac{\rho^r \,\rho^b}{\rho^2} \,\cos(\theta_i) \,f_i^{\mathrm{eq}}(\rho, 0), 
\end{align}
\noindent
where $\beta$ controls the interface thickness, with $\beta = 0.99$ used in all simulations to keep the interface as thin as possible, and $\theta_i$ is the angle between the color gradient $\mathbf{G}$ and the lattice connectivity vector $\mathbf{e}_i$.

\subsection{Model set-up}\label{section2.3}

\begin{figure*}[t]
\centering
\includegraphics[width=1\linewidth]{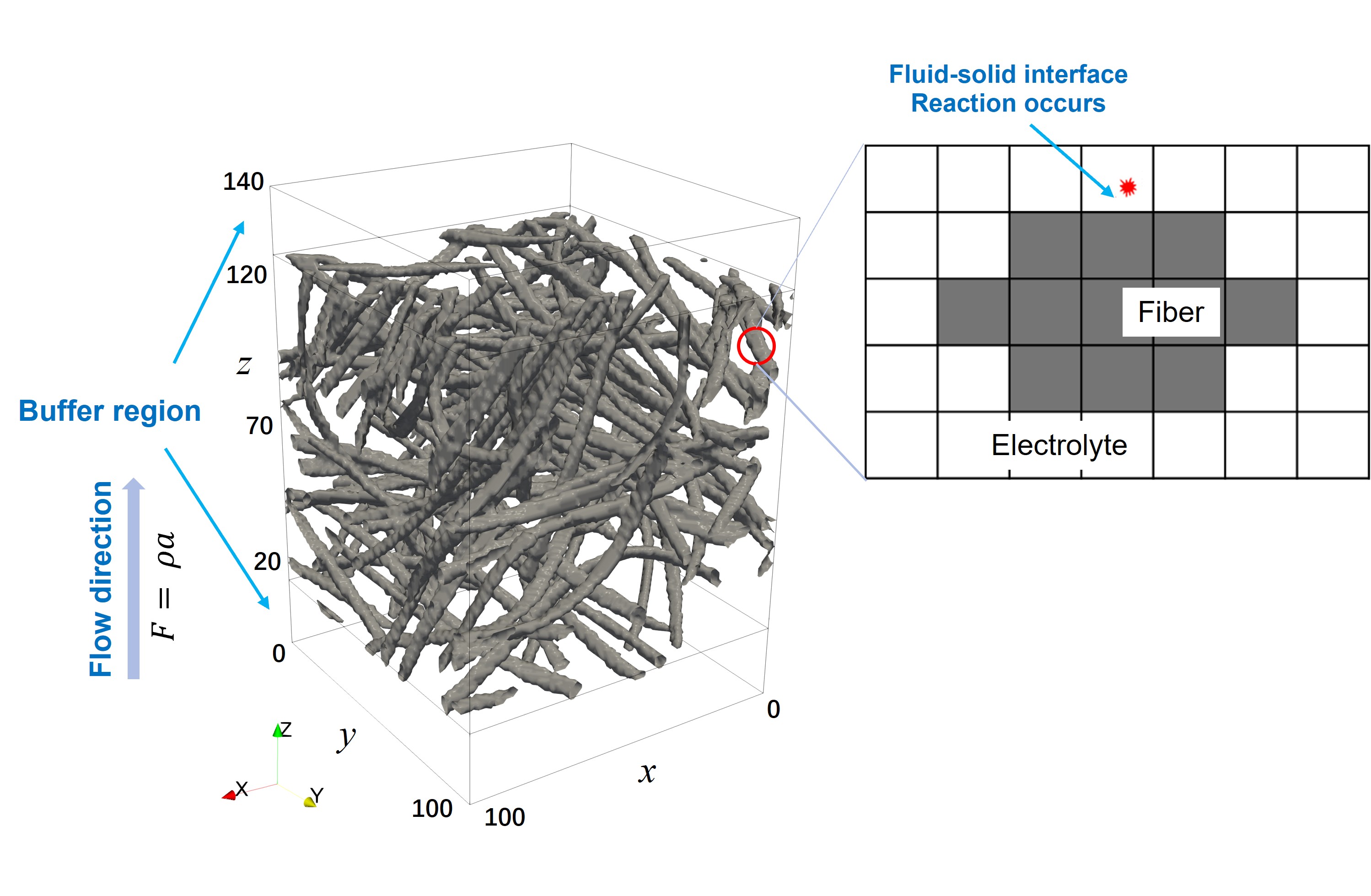} 
\caption{ Schematic diagram of the simulation geometry. }
\label{fig:modelsetup}
\end{figure*}

To simulate the bubble evolution over a longer time, 
we use a simulation domain of $100 \times 100 \times 100$ lattice sites (as shown in Fig.~\ref{fig:modelsetup}), which is sufficient to resolve the relevant pore-scale patterns while saving the computational cost. A domain-size sensitivity analysis and sample uncertainty are provided in Appendix~\ref{app:domain_size} and~\ref{app:uncertainty}.
The bubble generation procedure is introduced through reactiveconversion boundary conditions 
(Eq.~(\ref{eq:reaction})) at lattice nodes adjacent to the reactive boundary, converting all 19 fluid 
populations of one fluid species into another one governed by the reaction rate $k_r$ \cite{scheel2024},
\begin{equation}
\frac{\partial \rho^r}{\partial t} = - k_r\,\rho^r, \quad
\frac{\partial \rho^b}{\partial t} = k_r\,\rho^b.
\label{eq:reaction}
\end{equation}
\noindent
In this study, $k_r$ is treated as a nominal conversion rate prescribed to generate bubbles at the reactive boundary. It controls the bubble generation rate but is not linked to a predictive electrochemical current. Consequently, this study focuses on the investigation of kinematic bubble visualization and is limited to a qualitative pattern. 

To model the porous medium, the carbon fiber geometry is incorporated into the LBM domain, and a bounce-back boundary condition is applied at all solid--fluid interfaces to enforce a no-slip condition on the carbon fibers \cite{narvaez2013creeping}. The contact angle is prescribed as  60\textdegree, consistent with the hydrophilic behaviour of VRFB carbon felts in sulphuric electrolyte~\cite{ye2024vanadium, schilling2024investigating}, and is chosen to ensure sufficient wettability and numerical robustness with the current fictitious-density wetting scheme~\cite{latva2005static}. Because this scheme is reliable only over a moderate range of contact angles~\cite{chen2019inertial}, we do not investigate wettability dependence here, and our conclusions are limited to the hydrophilic regime.
The four faces parallel to the flow direction are set as periodic boundaries to emulate an interior representative sub-volume of the felt and to avoid unknown wall-wettability effects and artificial wall pinning. The inlet and outlet use the recoloring boundary to remove bubbles, as expressed in Eqs.~(\ref{eq:recolor1}) and (\ref{eq:recolor2}). This boundary is numerically robust for two-phase color-gradient LBM, reducing spurious interface reflections compared with standard open boundaries. The recoloring coefficients are determined based on the initial component densities $\rho_0^r = 1$ and $\rho_0^b = 1\times10^{-10}$. 
\begin{equation}
f_i^r\bigl({x}_\mathrm{boundary}, t\bigr)
= \frac{\rho_0^r}{\rho_0^r + \rho_0^b}\,
f_i\bigl({x}_\mathrm{boundary}, t\bigr),
\label{eq:recolor1}
\end{equation}
\begin{equation}
f_i^b\bigl({x}_\mathrm{boundary}, t\bigr)
= \frac{\rho_0^b}{\rho_0^r + \rho_0^b}\,
f_i\bigl({x}_\mathrm{boundary}, t\bigr).
\label{eq:recolor2}
\end{equation}
\noindent
where ${x}_\mathrm{boundary}$ represents a list of the lattice sites that belong to the inlet and outlet boundary. To eliminate the influence of complex geometry on the inlet and outlet boundary condition 
settings, we insert a buffer region with a thickness of 20 lattice sites at both ends \cite{narvaez2013creeping}. All saturation and bubble coverage statistics are sampled in the interior porous media region.

\subsection{Unit conversion}
To connect the parameters in the LBM simulation with real physical quantities, the lattice units must first be defined in terms of basic SI units, including the length scale $C_l$, the time scale $C_t$, and the mass scale $C_m$. These scales are chosen appropriately based on the physical properties such as surface tension, viscosity, and the resolution of the lattice,
\begin{equation}
C_t = \frac{C_l^2 \,\nu_{r,LB}}{\nu_{r,Phys}},
\end{equation}
\begin{equation}
C_m = \frac{\rho_{Phys}\,C_l^3}{\rho_{LB}}.
\end{equation}
Here $C_l$ is determined by the lattice resolution, $\nu_b$ is the kinematic viscosity of the blue fluid, and $\sigma$ is the surface tension. The subscripts Phys and LB represent physical and lattice Boltzmann units, respectively. Once these base units are defined, all other parameters in the LBM simulation can be computed using the corresponding dimensionless relations,
\begin{equation}
\sigma_{Phys} 
= \frac{\sigma_{LB}\,C_m}{C_t^2},
\end{equation}
\begin{equation}
P_{Phys}
= \frac{P_{\mathrm{LB}}\,C_m}{C_l\,C_t^2}.
\end{equation}
We assess the relative importance of body forces and interfacial tension using the Bond number (buoyancy-to-capillary) and the Capillary number (viscous-to-capillary). Taking the average pore size as the characteristic length, $L \approx 60\text{--}80~\mu\mathrm{m}$ for $\mathrm{CR}=25\%\text{--}75\%$ in present study, we obtain $\mathrm{Bo} = \dfrac{\Delta\rho\,g\,L^{2}}{\sigma}$ lies in the range of $10^{-5}$ to $10^{-3}$ and the capillary number $\mathrm{Ca} = \dfrac{\rho\,\nu\,u}{\sigma}$ is much smaller than $10^{-6}$. These ranges indicate a capillary-dominated regime in which buoyancy is negligible for pore-scale detachment, rise, and trapping. Consequently, we adopt an equal-density setting in the model ($\rho_{phys}^{r} = \rho_{phys}^{b} = 1.325 \times 10^3\,\mathrm{kg\,m^{-3}}$) as a kinematic simplification within this applicability domain. Previous work has applied and validated the color–gradient LBM for capillarity-driven phenomena, including droplet coalescence~\cite{scheel2023viscous}, evaporation~\cite{nath2025reaction}, and bubble growth and departure~\cite{scheel2024}. Furthermore, color-gradient LBM method have been used by different groups~\cite{Leclaire2017b,sedahmed2024wetting,zahid2025review} for bubble transport in porous media, which demonstrate its general applicability for mesoscale-macroscale phenomena. Then the pressure drop could be evaluated by the liquid density:
\begin{equation}
\Delta\,P
= \rho^{r}aL,
\end{equation}
\noindent
where $a$ is the acceleration, which is applied uniformly on both fluids at all fluid nodes, and $L$ is the distance between the inlet and outlet boundaries. For the sake of clarity, all simulation parameters discussed are presented in lattice unit.
\begin{table}[t]
\centering
\resizebox{\textwidth}{!}{
\begin{tabular}{l c r}
\hline
\textbf{Physicochemical parameters} & \textbf{Physical values} & \textbf{Lattice values} \\
\hline
Characteristic length, $L_0$ 
& $2.44 \times 10^{-6}\;\mathrm{m}$ 
& $1 \;\mathrm{l.u.}$ \\
Surface tension, $\sigma$ 
& $7.2 \times 10^{-2}\;\mathrm{N\,m}^{-1}$~\cite{qin2022measurement} 
& $0.069 \;\mathrm{l.u.}$ \\
Kinematic viscosity of H$_2$, $\nu_b$ 
& $1 \times 10^{-5}\;\mathrm{m}^2\mathrm{s}^{-1}$ 
& $0.228 \;\mathrm{l.u.}$ \\
Kinematic viscosity of electrolyte, $\nu_r$ 
& $4.38 \times 10^{-6}\;\mathrm{m}^2\mathrm{s}^{-1}$~\cite{SkyllasKazacos2016} 
& $0.1 \;\mathrm{l.u.}$ \\
Electrolyte density, $\rho^r$ 
& $1.325 \times 10^3\;\mathrm{Kg\,m}^{-3}$~\cite{SkyllasKazacos2016} 
& $1 \;\mathrm{l.u.}$ \\
Pressure, $P$ 
& $4.27\times 10^{5}\;\mathrm{Pa}$ 
& $1 \;\mathrm{l.u.}$ \\
Characteristic time, $t$ 
& $1.36 \times 10^{-2}\;\mathrm{s}$ 
& $1 \times 10^{5} \;\mathrm{l.u.}$ \\
\hline
\end{tabular}
}
\caption{\label{tab:table1}Parameters used in the present study.}
\end{table}

\section{Results and discussion}\label{section3}

\subsection{Channel flow}\label{section3.1}

\begin{figure*}[htbp]
\centering
\includegraphics[width=1\linewidth]{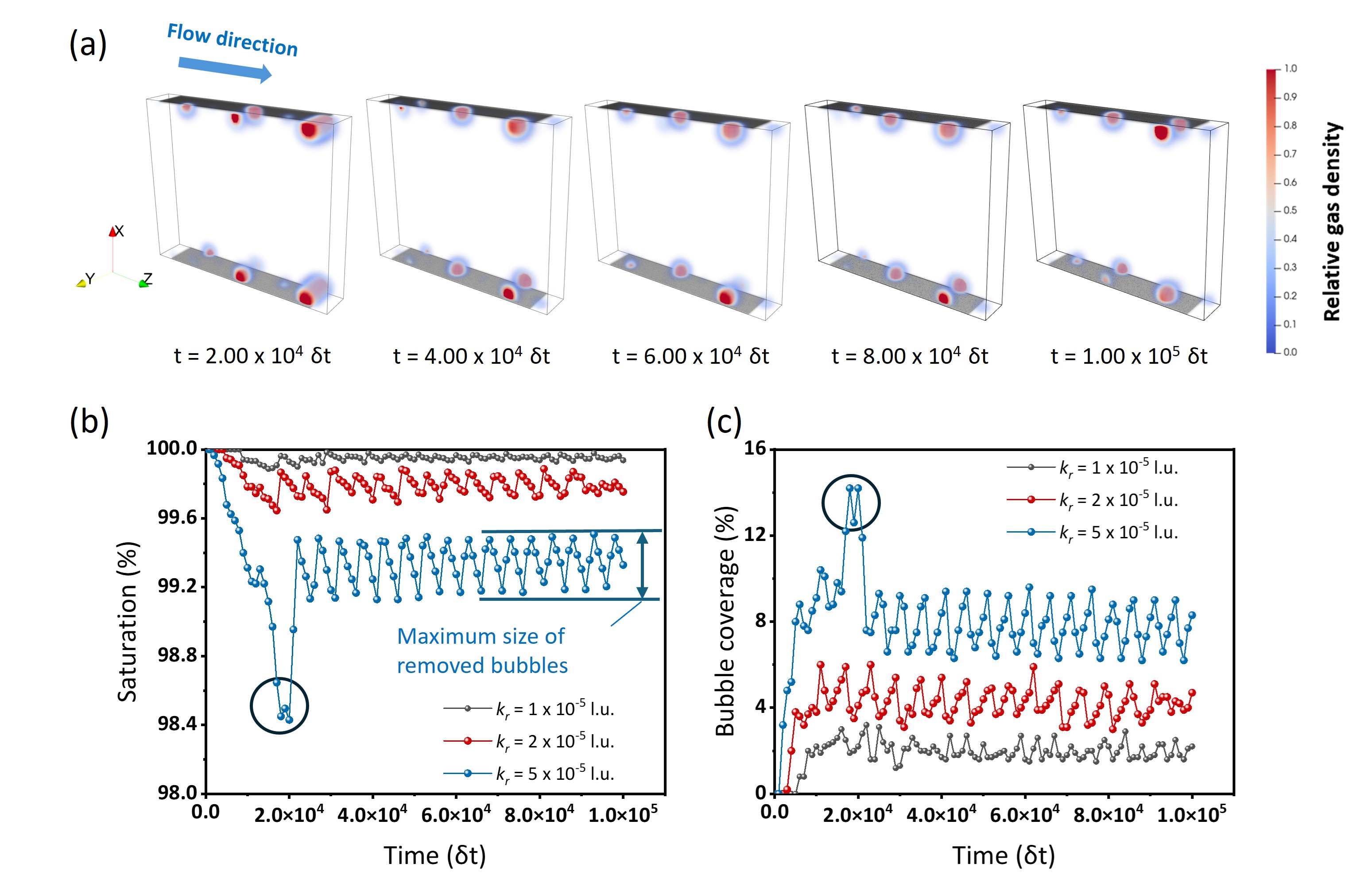} 
\caption{a) 3D visualization of bubble distribution in the electrolyte domain during the simulation. The color map represents the relative gas density; b) Electrolyte saturation curve plotted as a function of iteration steps for different reaction rates; c) Bubble coverage curve over time, also plotted for the same reaction rates.}
\label{fig:channel}
\end{figure*}

To gain a basic understanding of bubble evolution in complex geometries, we simulate a reactive flow in a simplified flat channel. The boundary conditions remain as described in Section~\ref{section2.3}, but the geometry is simplified to a channel ($50 \times 10 \times 60$ lattice sites), with two planar walls perpendicular to the x-axis serving as reaction walls. Initially, the computational domain is filled with a uniform electrolyte density of 1, which is then accelerated by a pressure drop ($\Delta\,P = 6\times10^{-4}\;\mathrm{l.u.}$). Fig.~\ref{fig:channel}(a) shows 3-D visualizations of bubble distribution in the channel. The removal of the gas phase is not continuous due to the formation of bubbles, which gradually grow. In order to have a qualitative analysis of bubble growth, we calculate the electrolyte saturation, shown in Fig.~\ref{fig:channel}(b), and the bubble coverage, shown in Fig.~\ref{fig:channel}(c), in the computational domain. The electrolyte saturation is defined as the ratio of the liquid volume to the total pore volume in the computational domain, while the bubble coverage is defined as the ratio of the bubble-solid contact area to the total surface area of the electrode. Both electrolyte saturation and bubble coverage exhibit quasi-periodic fluctuations once the flow reaches a quasi-steady state. Each fluctuation peak involves the coalescence, growth, and removal of a large bubble. However, at the maximum reaction rate case,  $kr = 5\times$ 10$^{-5} \;\mathrm{l.u.}$, a sudden jump is visible in Fig.~\ref{fig:channel}(b) and Fig.~\ref{fig:channel}(c) at  $t = 1.5\times 10^{4}$ $\delta t$. This behavior occurs during the early developing stage, where the flow rate is low and bubble growth is rapid. The limited geometric width of the channel restricts the bubble's growth, causing it to transition from hemispherical to a semi-cylindrical shape. Once the flow enters the steady state, the formation of semi-cylindrical bubbles no longer occurs. In this simple geometry, saturation and bubble coverage effectively capture the bubble evolution. Therefore, we continue using these two statistical variables in subsequent analyses.

\subsection{Effect of reaction rate}

The rate of the HER is influenced by various factors, including electrode materials, electrolyte properties, operating conditions, and the aging of the electrode. To anchor the numerical range used here, a roughly derivation is conducted based the parameters measured by Schweiss et al.~\cite{Schweiss2016}. It is found that the reaction rate $k_r$ should be on the order of $10^{-11}$ to $10^{-7}$. The bubble evolution under this $k_r$ value is too slow to be observed within our simulation timeframe, especially since we ignore the volume expansion effect. Therefore, for a qualitative analysis of how the HER rate influences bubble evolution, we accelerate the process by using reaction rate magnitudes ranging from $10^{-6}$ to $10^{-4}$. A more detailed derivation to anchor order-of-magnitude is provided in Appendix~\ref{app:kr}. Additionally, the pressure drop across the inlet and outlet is set to $1\times10^{-2} \;\mathrm{l.u.}$ to aid bubble transport within the simulation timeframe and the CR of the electrode is chosen as 25\%.

\begin{figure*}[t]
\centering
\includegraphics[width=1\linewidth]{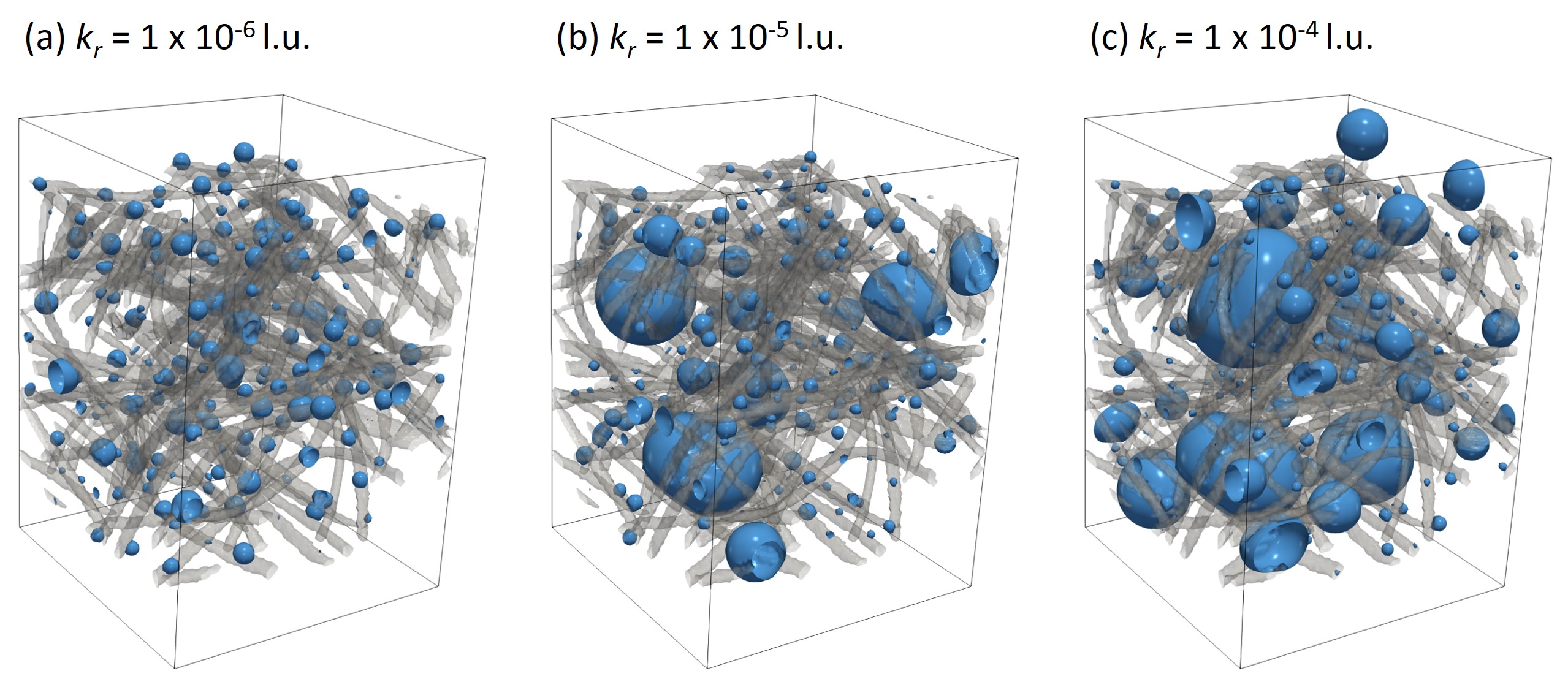} 
\caption{3-D visualization of the spatial bubble distribution at $t = 1.57\times 10^{5} \delta t$:
(a) $k_r = 1\times 10^{-6}\;\mathrm{l.u.}$; 
(b) $k_r = 1\times 10^{-5}\;\mathrm{l.u.}$; 
(c) $k_r = 1\times 10^{-4}\;\mathrm{l.u.}$.}
\label{fig:3dvisual}
\end{figure*}

\begin{figure*}[t]
\centering
\includegraphics[width=1\linewidth]{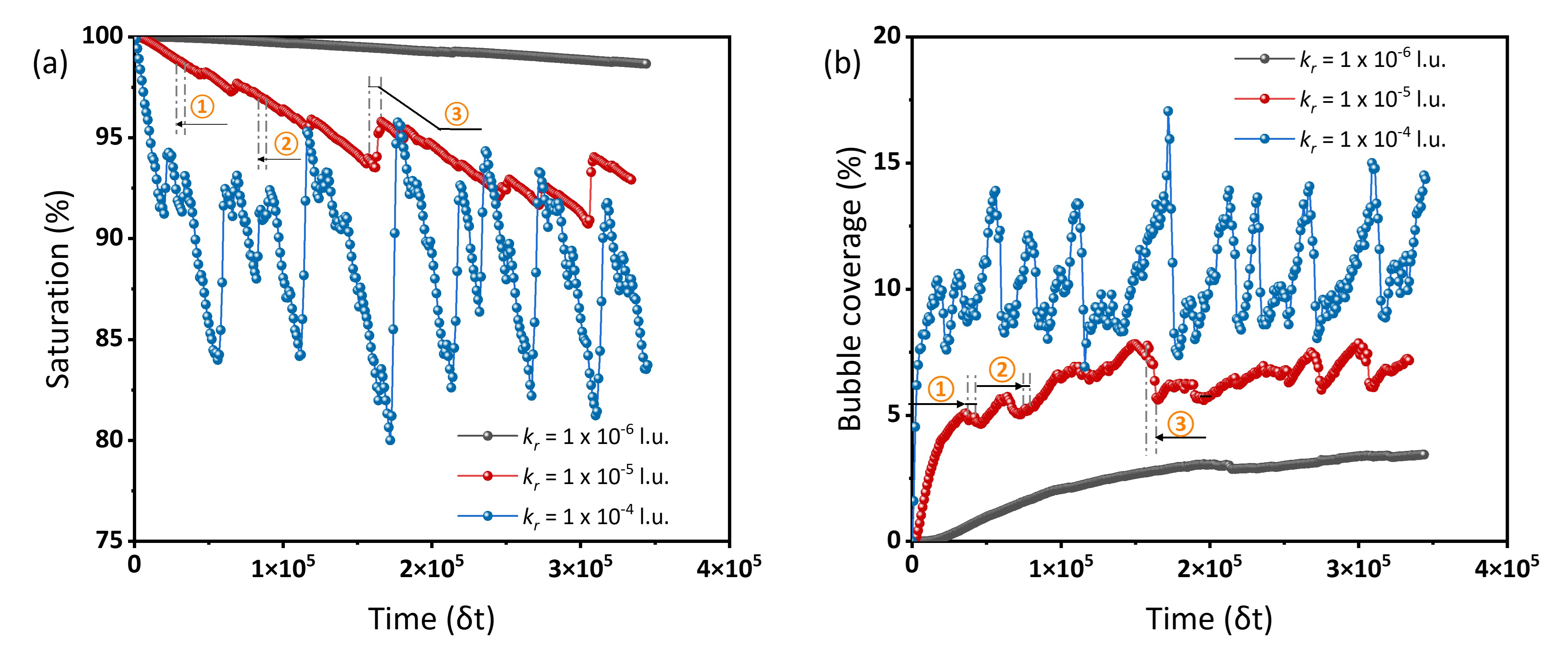} 
\caption{Variation of (a) electrolyte saturation and (b) bubble coverage over simulation time at different reaction rates.}
\label{fig:6}
\end{figure*}

Fig.~\ref{fig:3dvisual} illustrates the spatial distribution of bubbles at $t = 3.0\times 10^{5} \delta t$ for varying HER rates during reactive flow in the carbon felt, at the small reaction rate (Fig.~\ref{fig:3dvisual}(a)), bubbles grow slowly and remain mostly attached to the fiber surface. The movement of small bubbles toward each other and their coalescence are primarily driven by surface tension forces, which work to minimize the overall surface energy. At moderate reaction rates (Fig.~\ref{fig:3dvisual}(b)), bubbles grow more rapidly, and many of them detach from the fiber surface, moving under the influence of the flow. However, bubble movement is not always continuous; some bubbles coalesce during transport, becoming temporarily blocked by dense fiber areas. Once these coalesced bubbles reach a larger size, they are able to resume movement. As the reaction rate increases, bubbles grow larger and detach more rapidly from their fiber surface. This leads to a decrease in electrolyte saturation and a reduction in the effective reaction area, as larger bubbles occupy more space and obstruct the flow path. These effects are reflected in the electrolyte saturation and bubble coverage curves shown in Fig.~\ref{fig:6}, which display regular fluctuations, consistent with the trends observed in the channel flow study (section~\ref{section3.1}). Each sudden change in these statistical curves corresponds to the detachment and removal of large bubbles. However, the increased reaction rate also results in uneven gas release and greater obstruction to electrolyte transport, ultimately inhibiting the overall performance.

\begin{figure*}[t]
\centering
\includegraphics[width=1\linewidth]{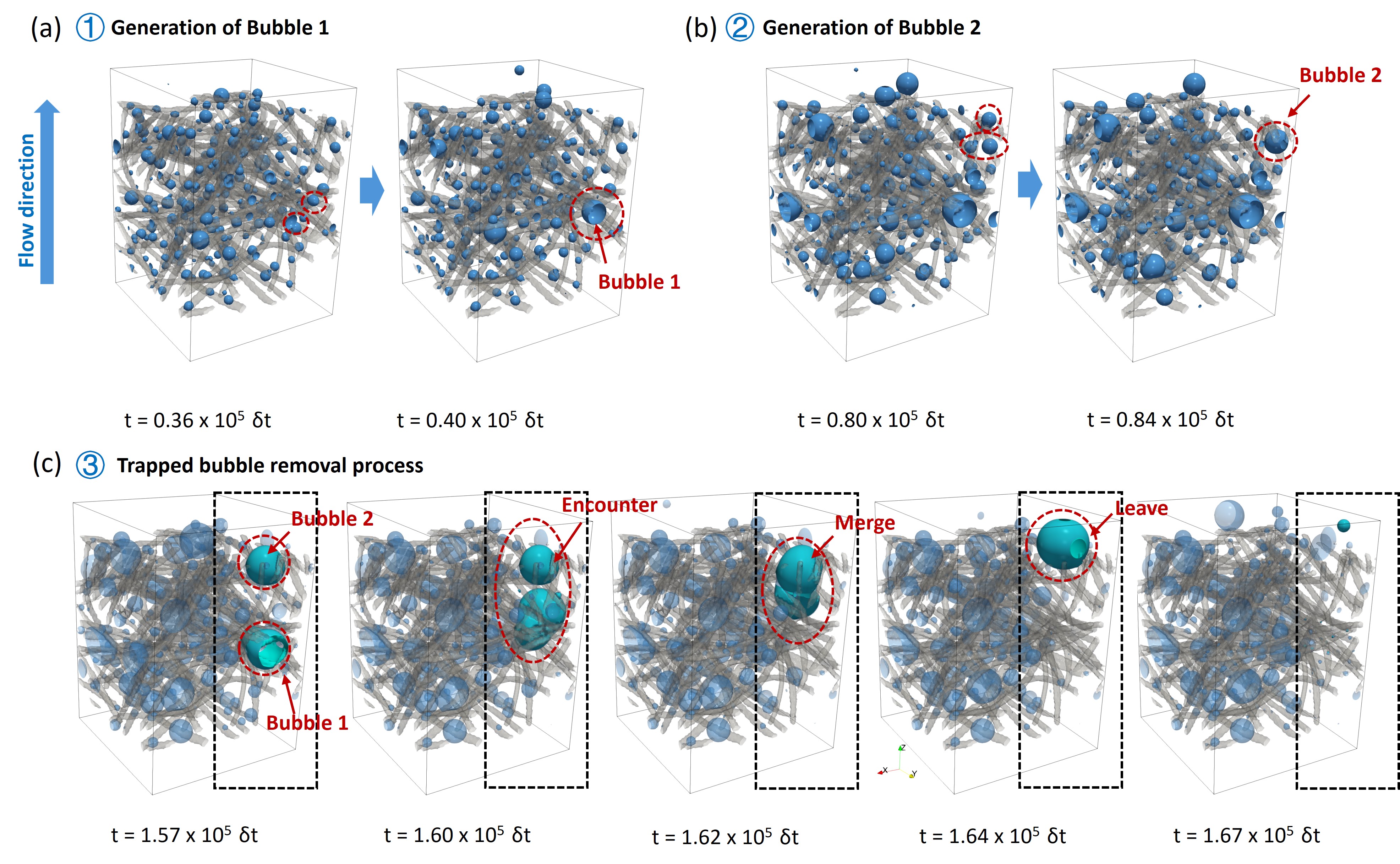} 
\caption{Bubble evolution process: (a) generation of bubble 1; (b) generation of bubble 2; (3) removal process of trapped bubble.}
\label{fig:bubbleremoval}
\end{figure*}

The removal of a trapped bubble within the porous electrode is further investigated at a reaction rate of $k_r = 1\times 10^{-5}\;\mathrm{l.u.}$, providing insights into bubble evolution dynamics. Fig.~\ref{fig:bubbleremoval} illustrates three stages of the bubble transport process highlighted also in Fig.~\ref{fig:6}: \ding{172} the coalescence of smaller bubbles to form Bubble 1, \ding{173} the coalescence of smaller bubbles form to Bubble 2, and \ding{174} the merging of Bubbles 1 and 2 and detachment of the resulting larger bubble. The tracked area is indicated by the black dashed box. Fig.~\ref{fig:bubbleremoval}(a) and~\ref{fig:bubbleremoval}(b) show the formation processes of Bubble 1 and Bubble 2, respectively. Both are formed by the coalescence of smaller “seed bubbles”, which are newly-formed small bubbles either recently detached from the carbon surface or still adhered around it. The movement of ‘seed bubbles’ is primarily governed by surface tension forces, including both solid–fluid (carbon surface) and fluid–fluid (bubble–bubble) interactions. These forces cause the bubbles to coalesce and influence their detachment from surfaces, which means their movement direction is not always aligned with the flow. Fig.~\ref{fig:bubbleremoval}(c) shows the bubble removal process. At $t = 1.57\times 10^{5} \delta t$, Bubble 1 and Bubble 2 are still observed to be trapped in the same position, with only their size increasing. Then Bubble 1 first breaks away from the fiber restraint and starts to move, eventually merging with Bubble 2 along its path and exiting the computational domain. From this perspective, bubbles are not trapped forever in so-called pinning sites. Each pinning site has different constraints on the bubble, depending on the surrounding microstructure. Once the bubble size exceeds its constraint limit, the bubble lifts off and moves away. Duan et~al.~\cite{duan2025investigating}, using synchrotron imaging, observed that bubbles with similar shape repeatedly pinned at the same electrode sites after each sweeping cycle. Figure 7 only depicts a few examples of the movement of bubbles in the electrode. The overall picture is composed of many similar processes causing fluctuations in the overall performance, which can be treated in a statistical manner.

\begin{figure*}[t]
\centering
\includegraphics[width=1\linewidth]{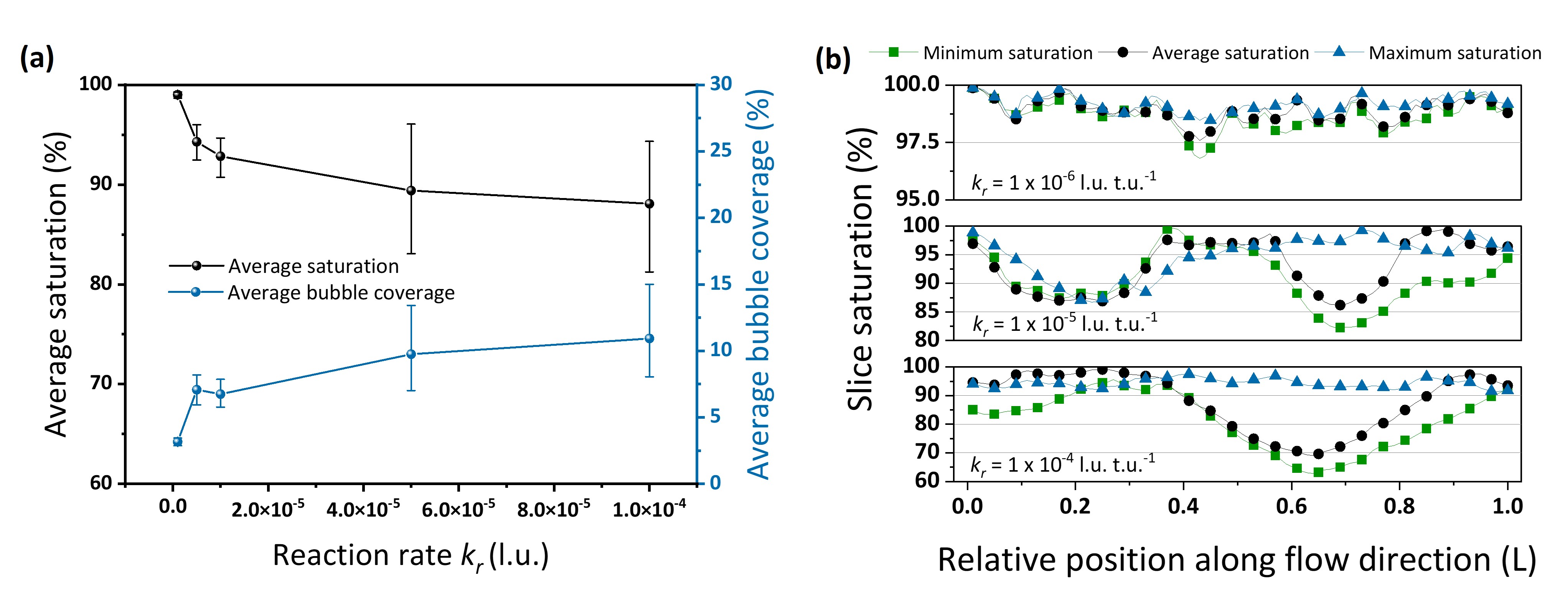} 
\caption{(a) Average electrolyte saturation and bubble’s coverage vary with reaction rate. Electrolyte slice saturation changes along the flow direction; (b) Electrolyte slice saturation changes along the flow direction.}
\label{fig:8}
\end{figure*}

Fig.~\ref{fig:8}(a) presents the average values of electrolyte saturation and bubble coverage at different reaction rates after 1.5 $\times$ 10$^5$ $\delta t$, indicated by solid dots. The error bars are computed by tracking the saturation values over time during the quasi-steady state. The top of the error bar is the maximum value observed, and the bottom is the minimum value observed during that period. The results show that when $k_r$ is below 1 $\times$ 10$^{-5}\;\mathrm{l.u.}$, saturation decreases sharply, while bubble coverage increases rapidly with rising $k_r$. However, when k$_r$ exceeds 1 $\times$ 10$^{-5}\;\mathrm{l.u.}$, both saturation and bubble coverage change more gradually. Fig.~\ref{fig:8}(b) shows variations in electrolyte slice saturation along the flow direction for different $k_r$, comparing their difference between the minimum, average, and maximum overall saturation levels. The horizontal axis in Fig.~\ref{fig:8}(b) is normalized from 0 to 1, with the unit being the total flow length $L$. At $k_r$ = 1 $\times$ 10$^{-6}\;\mathrm{l.u.}$, the difference in slice saturation curves is minimal due to the lack of bubble coalescence. However, at $k_r = 1 \times 10^{-5}$ and  $1\times$ 10$^{-4}\;\mathrm{l.u.}$, significant variations are observed in the slice saturation curves around the relative position of 0.4–0.8 $L$, where a distinct trough appears at the minimum saturation. This suggests the presence of a bubble pinning site in this region, aligning with the phenomenon observed in Fig.~\ref{fig:bubbleremoval}. Another pinning site could be observed in the region of 0.1–0.4 $L$. At $k_r = 1 \times 10^{-5}\;\mathrm{l.u.}$, the trapped bubbles remain largely unremoved, as indicated by the similar slice saturation values across average, maximum, and minimum saturation states. The lowest slice saturation in this region stays around 87\%, suggesting that the flow conditions are insufficient to dislodge these bubbles, resulting in a stable, persistent bubble distribution. In contrast, at $k_r = 1 \times 10^{-4}\;\mathrm{l.u.}$, a different behavior is observed. During the minimum saturation state, the lowest slice saturation in the region from 0.1 $L$ to 0.4 $L$ drops below 85\%, indicating the formation of more substantial gas voids. However, at the maximum saturation state, these trapped bubbles are effectively removed, suggesting that the higher reaction rate facilitates the periodic release of bubbles, which reduces localized gas accumulation. This observation aligns with Scheel et al.'s findings \cite{scheel2024}, which show that larger bubbles are removed more efficiently because coalescence facilitates their detachment.

\subsection{Effect of flow rate}

\begin{figure*}[t]
\centering
\includegraphics[width=1\linewidth]{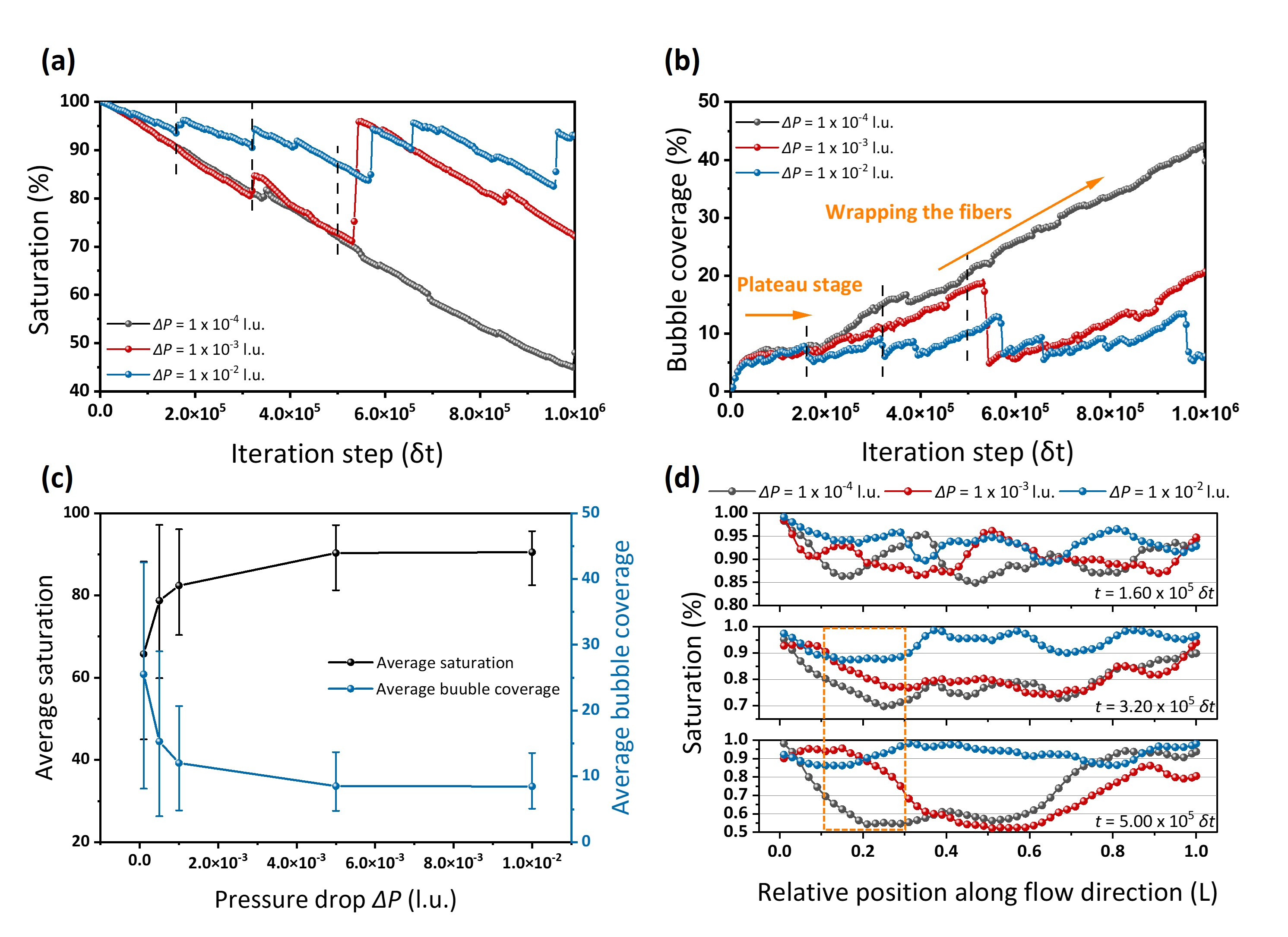} 
\caption{Variation of (a) electrolyte saturation and (b) bubble coverage over time at different values of the pressure drop; (c) average saturation and bubble coverage for different pressure drops; (d) slice saturation at $t = 1.60 \times 10^{5} \delta t$, $t = 3.20 \times 10^{5} \delta t$ and $t = 5.00 \times 10^{5} \delta t$.}
\label{fig:9}
\end{figure*}

The negative impact of bubbles on battery performance can be effectively reduced by adjusting the flow rate of the electrolyte. In our simulations, the flow rate is controlled by the pressure drop along flow direction. Here, 
$\Delta\,P = 1 \times 10^{-4}\;\mathrm{l.u.}$ is equivalent to applying a pressure drop of 42.7\,Pa between inlet and outlet of the porous media. The reaction rate $k_r$ is fixed as $1 \times 10^{-5}\;\mathrm{l.u.}$, and the CR is 25\%.

Fig.~\ref{fig:9} depicts the change of the electrolyte saturation and bubble coverage with time under different pressure drops. In order to more accurately analyze the bubble evolution, we extended the simulation duration to $t = 1.0 \times 10^{6} \delta t$ to better capture long-term bubble dynamics. At the lowest flow rate ($\Delta\,P = 1 \times 10^{-4}\;\mathrm{l.u.}$), the driving force is insufficient to promote efficient bubble removal process. As a result, bubble motion is significantly hindered, and gas accumulates within the porous structure. During the early stage of the simulation (before $t = 1.60 \times 10^{5} \delta t$), the saturation decreases monotonically in Fig.~\ref{fig:9}(a), while the bubble coverage exhibits a plateau in Fig.~\ref{fig:9}(b). Although the volume of bubbles in the electrode is growing, the bubble coverage does not vary much. This is due to bubbles that are not large enough to completely fill the local pores. Over time, as bubbles grow to a certain size, their interaction with the fibrous structure leads to deformation and wrapping around the fibers. This deformation initiates a phase of rapid growth in bubble coverage. Such behavior has also been observed in synchrotron imaging studies reported in the literature~\cite{Bevilacqua2019}. The wrapping effect facilitates coalescence with neighboring bubbles, leading to the formation of larger gas agglomerates. Once a bubble reaches sufficient size, the hydrodynamic and buoyancy forces acting on it can overcome the capillary constraints imposed by the porous structure, allowing it to detach and be carried away by the flow. This process repeats throughout the quasi-steady period, during which bubbles continuously nucleate, grow, coalesce, and detach in a dynamic equilibrium.

\begin{figure*}[t]
\centering
\includegraphics[width=1\linewidth]{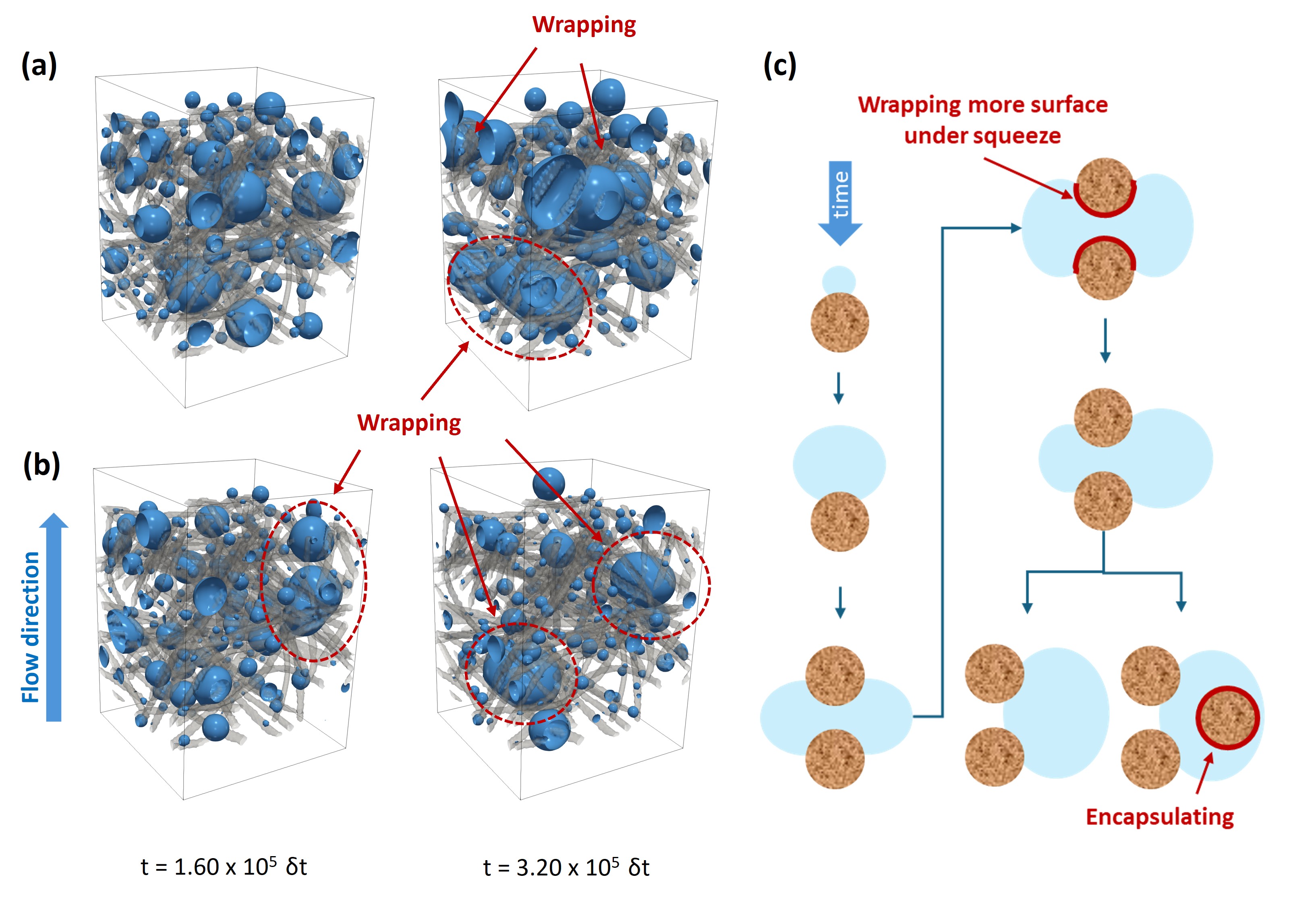} 
\caption{Spatial distribution of bubbles during the plateau ($t = 1.60 \times 10^{5} \delta t$) and rapid growth periods ($t = 3.20 \times 10^{5} \delta t$) of bubble coverage indicated in Figure 9b). (a) Distribution at $\Delta\,P = 1 \times 10^{-4}\;\mathrm{l.u.}$; (b) Distribution at $\Delta\,P = 1 \times 10^{-2}\;\mathrm{l.u.}$; (c) Schematic representation of the bubble wrapping process.}
\label{fig:10}
\end{figure*}

As the pressure drop increases, notable trends can be observed in bubble dynamics. From Fig.~\ref{fig:9}(a), it is evident that the maximum bubble size decreases, and the bubble removal frequency increases (i.e., the period shortens) with increasing flow rate. This indicates that higher pressure drops promote more efficient bubble detachment and transport. Meanwhile, Fig.~\ref{fig:9}(b) shows that the fluctuation amplitude of bubble coverage gradually decreases with increasing flow rate, suggesting a more uniform and stable two-phase flow regime at higher flow conditions. Furthermore, based on the average saturation and bubble coverage values calculated after $t = 2.0 \times 10^{5} \delta t$, a noticeable change in trend occurs at $\Delta\,P = 1 \times 10^{-3}\;\mathrm{l.u.}$ in Fig.~\ref{fig:9}(c). Below this threshold, average saturation increases sharply with pressure drop, while bubble coverage decreases significantly. Beyond this point, both saturation and bubble coverage vary more gradually with further increases in flow rate. From an energy-saving perspective, setting pressure drop to $\Delta\,P = 1 \times 10^{-3}\;\mathrm{l.u.}$ provides an optimal balance between bubble removal effectiveness and pumping energy input.

Fig.~\ref{fig:9}(d) shows the variation in slice saturation along the flow direction at three time steps, as marked by grey lines in Fig.~\ref{fig:9}(a) and Fig.~\ref{fig:9}(b). At $t = 1.60 \times 10^{5} \delta t$, all three cases exhibit relatively high bulk saturation values in Fig.~\ref{fig:9}(a), which are 90.0\%, 90.4\%, and 93.7\% respectively, and an overall even slice saturation distribution in Fig.~\ref{fig:9}(b). Among them, the case with $\Delta\,P = 1 \times 10^{-2}\;\mathrm{l.u.}$ shows the most stable and uniform distribution, with a fluctuation amplitude of approximately 5\%, compared to about 10\% in the other two cases. At $t = 3.20 \times 10^{5} \delta t$ and $t = 5.00 \times 10^{5} \delta t$, the slice saturation at $\Delta\,P = 1 \times 10^{-2}\;\mathrm{l.u.}$ remains relatively stable. In contrast, for $\Delta\,P = 1 \times 10^{-3}$ and $1 \times 10^{-4}\;\mathrm{l.u.}$, saturation decreases by around 15\% at $t = 3.20 \times 10^{5} \delta t$ and by about 25\% at $t = 5.00 \times 10^{5} \delta t$. In these lower-flow cases, bubbles grow large enough to wrap around the fibers, leading to a significant increase in bubble coverage, as shown in Fig.~\ref{fig:9}(b). The difference in the bubble coverage curve observed between $\Delta\,P = 1 \times 10^{-3}$ and $1 \times 10^{-4}\;\mathrm{l.u.}$ in Fig.~\ref{fig:9}(b) is attributed to the presence of a pinning site in the region between 0.1 $L$ and 0.3 $L$ along the flow direction as shown in Fig.~\ref{fig:9}(d). At the same reaction rate, a higher pressure drop facilitates the removal of trapped bubbles, improving saturation and reducing the number of bubbles wrapped around the fiber.

Fig.~\ref{fig:10} shows the spatial distribution of bubbles during the plateau ($t = 1.60 \times 10^{5} \delta t$) and rapid growth periods ($t = 3.20 \times 10^{5} \delta t$) of the bubble coverage as indicated in Fig.~\ref{fig:9}(b). Fig.~\ref{fig:10}(a) and~\ref{fig:10}(b) illustrate the distributions at $\Delta\,P = 1 \times 10^{-4}$ and $1 \times 10^{-2}\;\mathrm{l.u.}$, respectively. During the plateau period, bubble shapes are predominantly regular spherical caps in both cases, maintaining a stable contact with individual fibers. Due to the hydrophilic nature of the electrolyte, the interaction between bubbles and fibers at this stage does not result in significant wrapping, as the bubbles are limited to contacting single fibers and cannot spread further. In the rapid growth period ($t = 3.20 \times 10^{5} \delta t$), as bubbles grow larger, they begin to interact with multiple fibers, leading to wrapping under the constraints of the fibrous structure. At low flow rates, this wrapping is pronounced, with trapped large bubbles covering multiple fibers and significantly increasing surface coverage. However, at high flow rates, bubbles are more effectively removed, reducing the occurrence of fiber wrapping and maintaining better flow pathways within the electrode. To aid our understanding, Fig.~\ref{fig:10}(c) provides a schematic representation of the bubble-wrapping process. Initially, at smaller sizes, bubbles maintain a spherical cap shape and interact only with single fibers. As the bubbles grow larger, they extend to contact adjacent fibers, resulting in more wrapping due to spatial constraints within the porous structure. Under the influence of the hydrophobic surface, the bubbles gradually move toward the large pores. However, in areas with dense fibers and under the influence of bubble coalesce growth behavior, bubbles may encapsulate some fibers.

\subsection{Effect of compression ratios}

The porous media in VRFB systems are typically composed of carbon, which can undergo significant structural alterations under different levels of compression or material stress. As such, understanding the effects of compression on the pore structure of carbon and its porosity is critical for optimizing bubble transport. This section investigates the effects of different microstructures on bubble transport. To accelerate the analysis of bubble evolution in porous media, the reaction rate $k_r$ is fixed as $1 \times 10^{-5}\;\mathrm{l.u.}$ and the pressure drop across the inlet and outlet is set to $1\times10^{-2} \;\mathrm{l.u.}$.

\begin{figure}[htbp]
\centering
\includegraphics[width=1\linewidth]{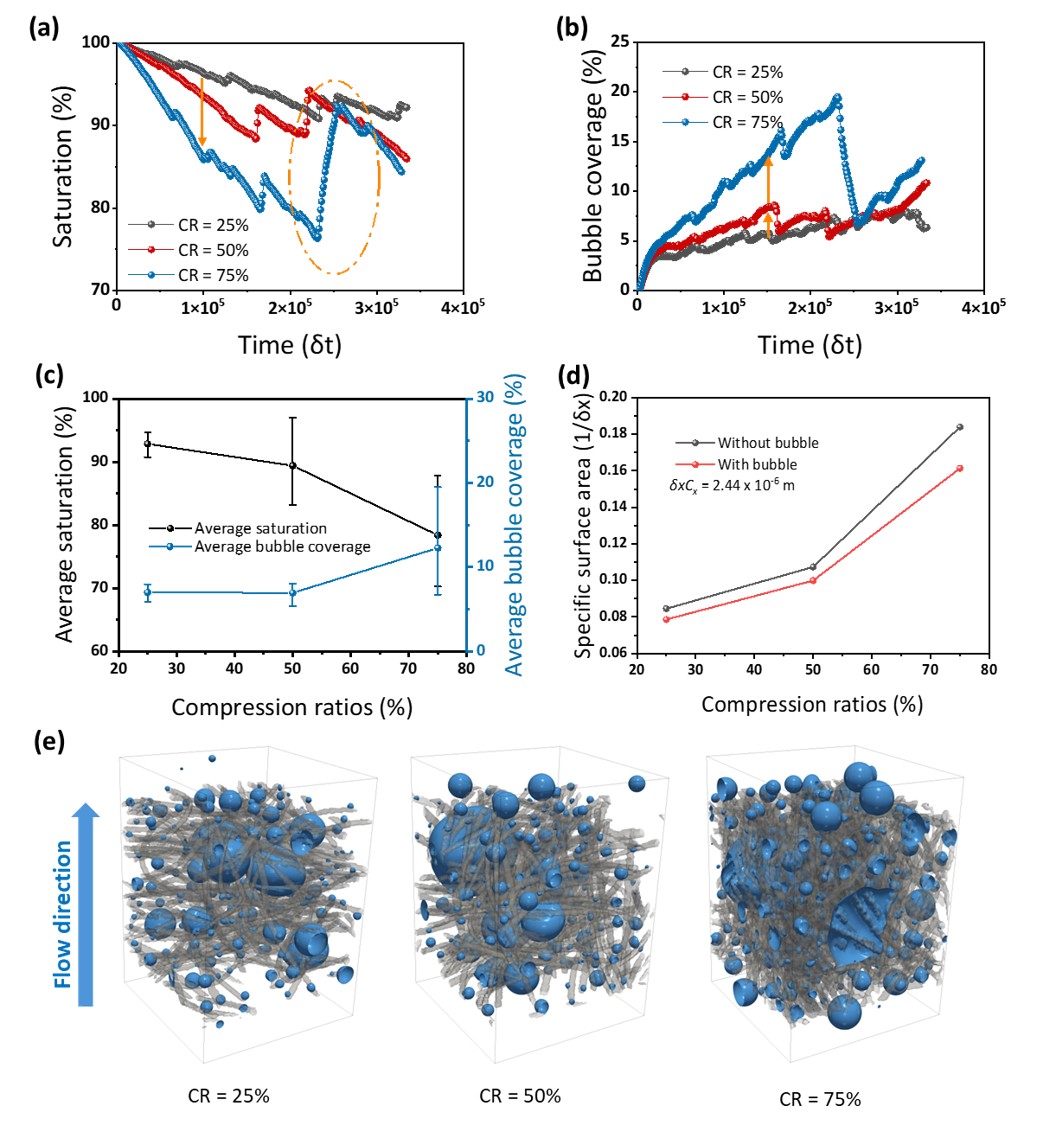} 
\caption{Variation of (a) electrolyte saturation and (b) bubble coverage over time at different compression ratios; (c) average saturation and bubble coverage for different compression ratios; (d) specific surface area for different compressin ratios (with and without bubbles); (e) visualization of bubble distribution at $t = 2.0 \times 10^{5} \delta t$.}
\label{fig:11}
\end{figure}

From the slope of the saturation curve in Fig. 11(a) at the early stage of the simulation, it can be seen that higher CRs lead to an increased gas generation rate. This trend is attributed to the enhanced spatial density of catalytic sites in the compressed carbon felt, which potentially increases spatial catalytic efficiency. However, the amplitude of the saturation curve also indicates that larger bubbles tend to be trapped under higher CRs. These larger trapped bubbles occupy more pore space, hinder the transport of reactants and reduce the effective reaction area, ultimately leading to a decrease in catalytic efficiency. The coverage curve in Fig. 11(b) further highlights that bubble coverage increases slightly as the CR increases from 25\% to 50\% but grows significantly when the CR increases from 50\% to 75\%. This trend can be observed intuitively in the average value curves shown in Fig. 11(c). The sharp rise in bubble coverage is attributed to the larger bubble sizes and the denser fiber network structure under higher CRs, especially when increasing the CR from 50\% to 75\%. Fig.~\ref{fig:11}(d) show the specific surface area, which is defined as the geometric surface to volume ratio of the carbon felt, varies with compression ratio. The black curve represents an ideal, fully wetted case, while the red curve treats bubble-covered regions as unavailable. With increasing CR, the specific surface area rises markedly (densification of fibers), and the red curve shows more deviation from the black one due to more severe bubble shielding. This trend is consistent with the behavior observed in Fig.~\ref{fig:11}(a). Lastly, Fig.~\ref{fig:11}(e) illustrates the spatial bubble distribution for varying CRs, showing that trapped big bubbles in porous domain grow progressively larger with increasing CR, further exacerbating the obstruction of electrolyte flow.

Moreover, excessive compression can mechanically deform the carbon fibers, reduce pore connectivity, and compromise the structural integrity of the electrode over long-term cycling. Prior studies have shown that high CRs may exacerbate capacity decay by accelerating vanadium ion crossover and increasing hydraulic resistance~\cite{jeong2023enhancing, hsieh2019effect, bromberger2014model, wang2023numerical}. On the other hand, compression is necessary to ensure good contact between fibers, reduce electronic resistance, and prevent electrolyte leakage. Li et al.~\cite{Li2020b} and Xiao et al.~\cite{xiao2020pore} demonstrated that increasing CR can reduce diffusivity and permeability, but improve electronic conductivity, highlighting the trade-off between mass transport and charge conduction in porous electrodes. This trade-off between catalytic accessibility and transport limitation has also been emphasized in the design of structured carbon electrodes aiming to maximize spatial catalysis, where interstitial space utilization must be balanced against pore blockage and flow constraints~\cite{sheng2018robust}. Given these competing effects, including enhanced reaction site density, impeded mass transport, local bubble trapping, mechanical degradation risks, and electrical benefits, it is clear that compression must be carefully optimized. Based on our results and existing literature, a moderate CR in the range of 25\% to 50\% is recommended to strike a balance between catalytic accessibility, transport efficiency, and mechanical robustness in VRFB applications.

\section{Conclusion}\label{section4}

This study empolys a color-gradient lattice Boltzmann model, combined with X-ray synchrotron images, to investigate bubble formation and transport in VRFBs, capturing the interplay between dynamic bubbles and electrode microstructures. Our findings show that bubble evolution can significantly obstruct electrolyte flow, hightlighting the needs to balance reaction rates, flow rates, and electrode microstructures to effectively manage bubble formation and transport in VRFBs. 

Within the capillary-dominated pore-scale regime (\( \mathrm{Ca} \), \( \mathrm{Bo}\ll 1 \)), we report qualitative, dimensionless trends. Increasing the nominal reaction rate promotes bubble growth and detachment, while insufficient flow rates cause persistent bubble trapping. Higher compression ratio, though densifies the felt (raising geometric specific surface area), while also exacerbates bubble retention.

Although this work focuses on VRFBs, the mechanisms identified—bubble nucleation, growth, transport, and interactions with fibrous porous structures—are relevant to a broad class of electrochemical systems employing carbon fiber electrodes. While demonstrated here with a typical commercial carbon felt, the modeling approach is broadly applicable to other porous electrode architectures, including gas diffusion layers in proton exchange membrane fuel cells and electrolyzers.

The present framework is a dynamic bubble investigation in porous electrode. It does not couple charge conservation, species transport, or dissolved-gas physics, and it uses equal-density, immiscible phases to isolate capillarity. Building on this, future work will introduce a minimal electrochemical coupling to relate bubble coverage to performance, and adopt a more accurate wetting boundary for a systematic study of electrode-surface properties.

\section{Acknowledgments}\label{acknowledgments}
The authors express their sincere gratitude to SGL Carbon for generously providing the SIGRACELL\textsuperscript{\textregistered} carbon felts used in this research. K. D. acknowledges the financial support received from the China Scholarship Council under grant number 202106950013. The authors extend their thanks to the KIT light source for granting access to their beamlines and also express their appreciation to the Institute for Beam Physics and Technology (IBPT) for their dedicated operation of the storage ring, the Karlsruhe Research Accelerator (KARA). Q. X. and J. H. acknowledge the Deutsche Forschungsgemeinschaft (DFG, German Research Foundation)—Project No. 431791331—SFB 1452 and the Federal Ministry of Education and Research (Germany)—project H2Giga/AEM-Direkt (Grant number 03HY103HF) for funding and the Gauss Centre for Supercomputing e.V. (\url{www.gauss-centre.eu}) for funding this project by providing computing time through the John von Neumann Institute for Computing (NIC) on the GCS Supercomputer JUWELS at Jülich Supercomputing Centre (JSC). The underlying plotted data are available at \href{https://doi.org/10.5281/zenodo.17550793}{Zenodo (10.5281/zenodo.17550793)}.

\appendix

\makeatletter
\renewcommand\thesubsection{\Alph{section}.\arabic{subsection}} 
\makeatother
\numberwithin{equation}{subsection}
\numberwithin{figure}{subsection}
\numberwithin{table}{subsection}
\section{}

\subsection{Model validation}\label{app:validation}
\begin{figure}[htbp]
\centering
\includegraphics[width=1\linewidth]{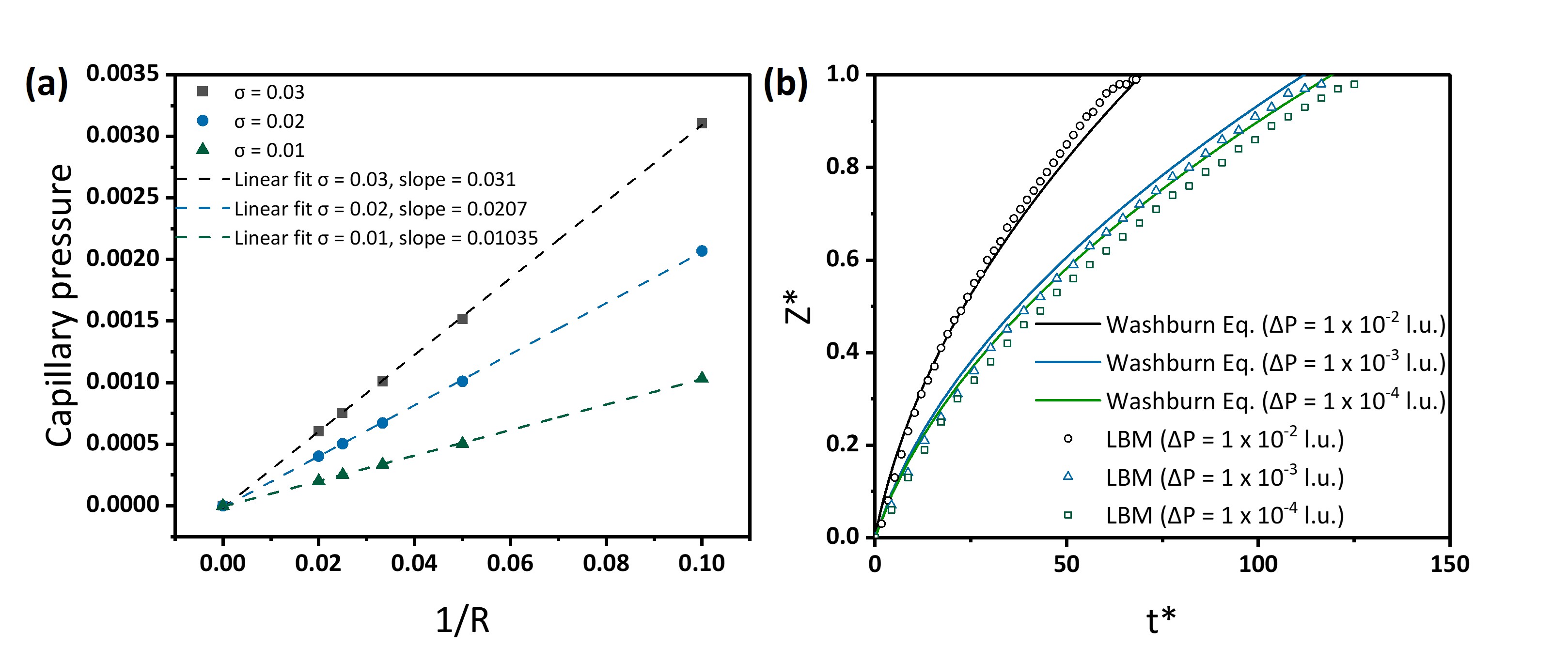} 
\caption{Benchmarks: (a) Laplace law test results for three surface tension values; (b) comparison between the Washburn prediction and LBM simulations for three imposed pressure gradients.}
\label{fig:validation}
\end{figure}

In this study, the color-gradient model is validated with two standard benchmarks: Laplace’s law and the Washburn capillary displacement. Figure~\ref{fig:validation}(a) shows the Laplace test for three surface–tension values. The capillary pressure varies linearly with  1/R and the fitted slopes match the prescribed $\sigma$, confirming accurate interfacial tension. The lattice resolution used here is deliberately coarse so this test verifies performance under the intended mesh. Figure~\ref{fig:validation}(b) compares the LBM results with the non-dimensional Washburn prediction for three imposed pressure gradients. All cases captured the underlying physics properly, where the fluids interface moved more rapidly at higher pressures, and it quickly moved at first, then slowed down as it progressed through the tube. Together, these benchmarks demonstrate that the model reliably captures capillarity-driven displacement in small pores at coarse resolution.

\subsection{Domain-size sensitivity}\label{app:domain_size}
\begin{figure}[htbp]
\centering
\includegraphics[width=0.5\linewidth]{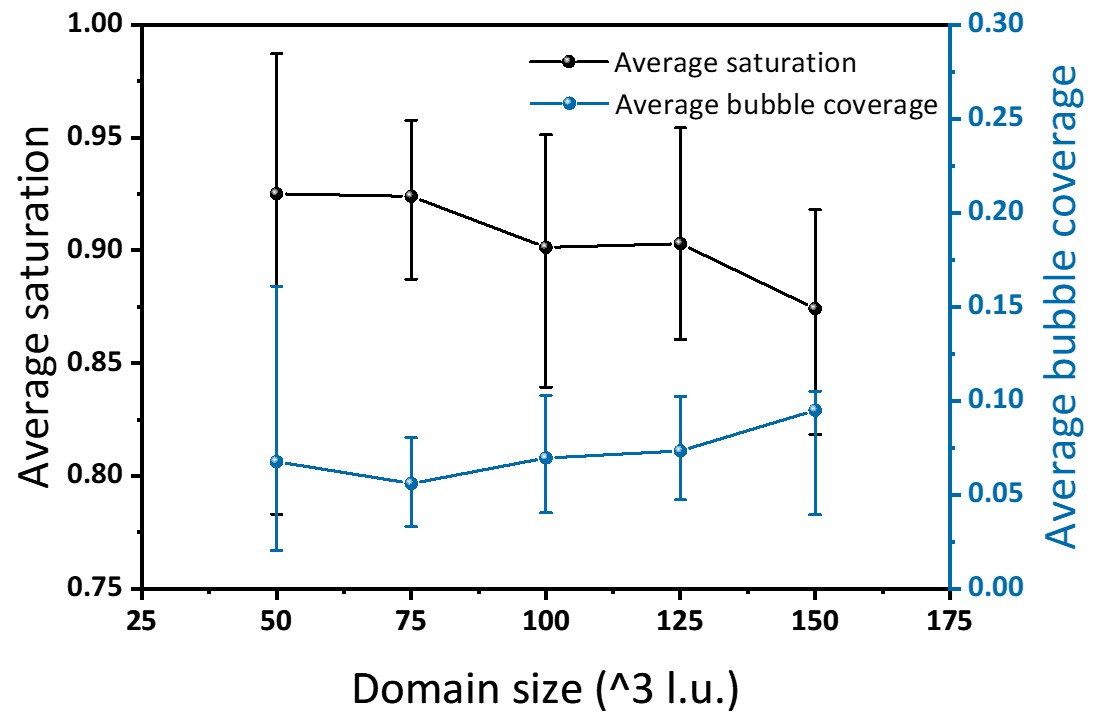} 
\caption{Average saturation varies with domain size.}
\label{fig:domain_size}
\end{figure}

Figure~\ref{fig:domain_size} reports the average electrolyte saturation varies as the porous domain size increases (in lattice units), with error bars indicating maximum and minimum transient values. The 50$^3$ l.u. case exhibits noticeably larger error bars. We attribute this to the interaction of bubbles with the periodic sidewalls in a small domain, since truncated fibers at the periodic cuts can create artificial pinning sites. As the domain size increase, average electrolyte saturation decreases mildly, consistent with longer internal pathways and reduced boundary influence. For sizes $\ge 100^3$ l.u., the average saturation and its variability stabilize; the change from $100^3$ to $150^3$ l.u.\ remains within 5-10\%. We therefore adopt $100^3$ l.u. as the baseline size, it is large enough to capture the relevant pore-scale patterns while saving the computational cost.   

\subsection{Sample uncertainty}\label{app:uncertainty}
\begin{figure}[h]
\centering
    \includegraphics[width=.5\linewidth]{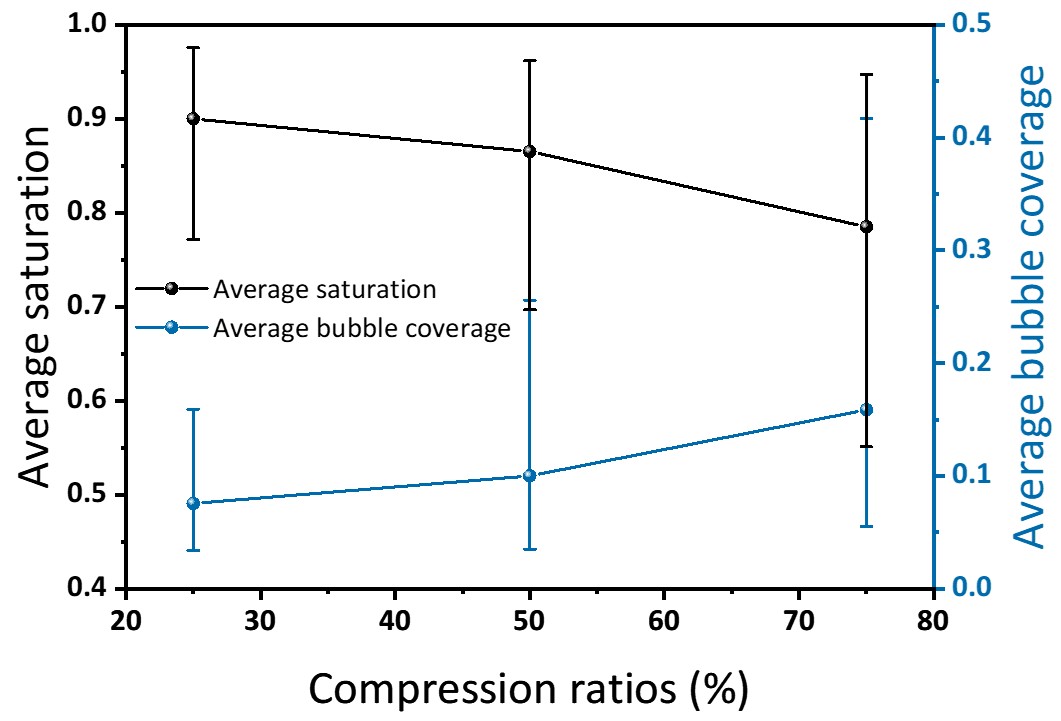}
    \caption{Ensemble average electrolyte saturation and bubble coverage for different compression ratios.}
\label{fig:ensemble_CR}
\end{figure}

To provide a baseline sense of sampling reliability without an exhaustive ensemble, we simulate five non-overlapping interior subvolumes for each compression ratio (CR = 25\%, 50\%, 75\%). Each subvolume is $100\times100\times100$ lattice units and uses the same boundary and source settings as in the main text. Statistics are sampled only in the interior porous region. We report the ensemble mean and standard deviation across the five subvolumes for average electrolyte saturation and bubble coverage in Table~\ref{tab:uncertainty}. It could be found that all standard deviations are below 5\%. The curves in Figure~\ref{fig:ensemble_CR} shows the average of the five cases and the error bars span the minimum–maximum range across the five cases, which is not the standard deviation. The trends are consistent with that in the main text. As compression ratio increases, the average saturation decreases while the average bubble coverage increases, and the error bars widen with compression, indicating larger size of trapped bubbles at higher CR.

\begin{table}[h]
\centering
\caption{Ensemble statistics across five interior subvolumes per compression ratio.}
\label{tab:uncertainty}
\begin{tabular}{lccc}
\hline
Variable & Condition & Mean & Standard deviation \\
\hline
Electrolyte saturation & CR = 25\% & 0.9000 & 0.0204 \\
                       & CR = 50\% & 0.8652 & 0.0240 \\
                       & CR = 75\% & 0.7852 & 0.0339 \\
Bubble coverage        & CR = 25\% & 0.0757 & 0.0060 \\
                       & CR = 50\% & 0.1001 & 0.0249 \\
                       & CR = 75\% & 0.1587 & 0.0430 \\
\hline
\end{tabular}
\end{table}

\subsection{Derivation of the nominal reaction rate}\label{app:kr}

This appendix provides a dimensional anchor only for the nominal boundary conversion rate \(k_r\) used in Eq.~\ref{eq:reaction}. It maps representative electrochemical currents to an order-of-magnitude source so that trends in the discussion section can be expressed via non-dimensional controls. It is not intended for predictive HER kinetics.

For the HER, a cathodic kinetic expression gives the local current density
\begin{equation}
j_{\mathrm{H_2/H^+}}
= a_v\, j_0^{\mathrm{H_2/H^+}}
  \exp\!\left(\frac{-2(1-\beta)F\,\eta_{\mathrm{H_2/H^+}}}{RT}\right)
\label{eq:appA_BV}
\end{equation}
where \(a_v\) is the geometric specific surface area, \(j_0\) is the exchange current density (e.g.\ \(6.6\times10^{-5}\) to \(2.7\times10^{-1}\,\mathrm{mA\,cm^{-2}}\) as reported by Schweiss et al.~\cite{Schweiss2016}), \(\beta\) is the transfer coefficient, \(F\) is Faraday’s constant, \(R\) is the general gas constant, \(T\) is temperature, and \(\eta_{\mathrm{H_2/H^+}}\) is the overpotential. 
In present work, we assume a hydrogen-ion concentration of \(4\,\mathrm{M}\) and an applied potential of \(-0.26\,\mathrm{V}\). The nominal reaction rate \(k_r\) is then estimated as
\begin{equation}
k_r = \frac{ j_{\mathrm{H_2/H^+}} M_{\mathrm{H_2}} }{ 2F \rho_{\mathrm{H_2}} } C_t
\end{equation}
where \(M_{\mathrm{H_2}}\) is the molar mass, $rho_{\mathrm{H_2}}$  is hydrogen density and $C_t$ LBM time scale conversion factor. This back-of-the-envelope estimate indicates that \(k_r\) in lattice units is very small (typically \(10^{-11}\)–\(10^{-7}\)), which would make bubble generation too slow to observe within feasible runtimes under our equal-density, incompressible setting. To accelerate bubble generation for qualitative visualization, we apply a factor based on the liquid-to-gas density ratio (order \(10^{5}\)). It is thus reasonable to explore \(k_r\) in the range \(10^{-6}\) to \(10^{-2}\) (l.u.) in the main text, while interpreting results through non-dimensional trends rather than quantitative kinetics.









\bibliographystyle{elsarticle-num} 
\biboptions{sort&compress} 
\bibliography{reference}

@article{Guney2017,
  title={Classification and assessment of energy storage systems},
  author={Guney, Mukrimin Sevket and Tepe, Yalcin},
  journal={Renew. {S}ustain. {E}nergy {R}ev.},
  volume={75},
  pages={1187--1197},
  doi={10.1016/j.rser.2016.11.102},
  year={2017}
}

@article{Lourenssen2019,
  title={Vanadium redox flow batteries: {A} comprehensive review},
  author={Lourenssen, Kyle and Williams, James and Ahmadpour, Faraz and Clemmer, Ryan and Tasnim, Syeda},
  journal={J. {E}nergy {S}torage},
  volume={25},
  pages={100844},
  doi={10.1016/j.est.2019.100844},
  year={2019}
}

@article{Ryan2019,
  title={Mesoscale modeling in electrochemical devices—{A} critical perspective},
  author={Ryan, Emily M and Mukherjee, Partha P},
  journal={Prog. {E}nergy {C}ombust. {S}ci.},
  volume={71},
  pages={118--142},
  doi={10.1016/j.pecs.2018.11.002},
  year={2019}
}

@article{Kear2012,
  author={Kear, Gareth and Shah, Akeel A and Walsh, Frank C},
  title     = {Development of the all-vanadium redox flow battery for energy storage: {A} review of technological, financial and policy aspects},
  journal={Int. {J}. {E}nergy {R}es.},
  volume={36},
  number={11},
  year={2012},
  doi={10.1002/er.1863},
  pages     = {1105--1120}
}

@article{Koble2021,
  title={Synchrotron {X}-{R}ay radiography of vanadium redox flow batteries--{T}ime and spatial resolved electrolyte flow in porous carbon electrodes},
  author={K{\"o}ble, Kerstin and Eifert, L{\'a}szl{\'o} and Bevilacqua, Nico and Fahy, Kieran F and Bazylak, Aimy and Zeis, Roswitha},
  journal={J. {P}ower {S}ources},
  volume={492},
  pages={229660},
  doi={10.1016/j.jpowsour.2021.229660},
  year={2021}
}

@article{duan2025investigating,
  title={Investigating Bubble Formation and Evolution in Vanadium Redox Flow Batteries via Synchrotron {X}-Ray Imaging},
  author  = {Kangjun Duan and Kerstin K{\"o}ble and Alexey Ershov and Monja Schilling and Alexander Rampf and Angelica Cecilia and Tom{\'a}{\v{s}} Farag{\'o} and Marcus Zuber and Tilo Baumbach and Pang-chieh Sui and Roswitha Zeis},
  journal={Chem{S}us{C}hem},
  eid={e2500282},
  volume={00},
  doi={10.1002/cssc.202500282},
  year={2025}
}

@article{Bevilacqua2019,
  title={Visualization of electrolyte flow in vanadium redox flow batteries using synchrotron {X}-ray radiography and tomography--{I}mpact of electrolyte species and electrode compression},
  author={Bevilacqua, Nico and Eifert, L{\'a}szl{\'o} and Banerjee, Rupak and K{\"o}ble, Kerstin and Farag{\'o}, Tom{\'a}{\v{s}} and Zuber, Marcus and Bazylak, Aimy and Zeis, Roswitha},
  journal={J. {P}ower {S}ources},
  volume={439},
  pages={227071},
  doi={10.1016/j.jpowsour.2019.227071},
  year={2019}
}

@article{Huang2022,
  title={Comprehensive analysis of critical issues in all-vanadium redox flow battery},
  author={Huang, Zebo and Mu, Anle and Wu, Longxing and Yang, Bin and Qian, Ye and Wang, Jiahui},
  journal={{ACS} {S}ustain. {C}hem. {E}ng.},
  volume={10},
  number={24},
  pages={7786--7810},
  doi={10.1021/acssuschemeng.2c01372},
  year={2022}
}

@article{Ye2017,
  title={Thermally induced evolution of dissolved gas in water flowing through a carbon felt sample},
  author={Ye, Qiang and Shan, Tian-Xiang and Cheng, Ping},
  journal={Int. {J}. {H}eat Mass {T}ransf.},
  volume={108},
  pages={2451--2461},
  doi={10.1016/j.ijheatmasstransfer.2017.01.097},
  year={2017}
}

@article{Ye2020,
  title={Effects of wettability and flow direction on gas retention and flow resistance of water flowing through carbon felts with thermally induced gas evolutions},
  author={Ye, Qiang and Zhang, Yu-Jia and Cheng, Ping and Shao, Zhigang},
  journal={Int. {J}. {H}eat Mass {T}ransf.},
  volume={156},
  pages={119911},
  doi={10.1016/j.ijheatmasstransfer.2020.119911},
  year={2020}
}

@article{Zhao2019,
  title={Gas bubbles in electrochemical gas evolution reactions},
  author={Zhao, Xu and Ren, Hang and Luo, Long},
  journal={Langmuir},
  volume={35},
  number={16},
  pages={5392--5408},
  doi={10.1021/acs.langmuir.9b00119},
  year={2019}
}

@article{AlNajjar2024,
   author={Al Najjar, Taher and Omran, Mostafa M and Allam, Nageh K and El Sawy, Ehab N},
  title     = {Tungsten oxide nanostructures for all-vanadium redox flow battery: {E}nhancing the {V}({II})/{V}({III}) reaction and inhibiting {H$_2$} evolution},
  journal={J. {E}nergy {S}torage},
  volume={79},
  pages={110123},
  doi={10.1016/j.est.2023.110123},
  year={2024}
}

@article{Wen2023,
  title={Bismuth concentration influenced competition between electrochemical reactions in the all-vanadium redox flow battery},
  author={Wen, Yue and Neville, Tobias P and Sobrido, Ana Jorge and Shearing, Paul R and Brett, Dan JL and Jervis, Rhodri},
  journal={J. {P}ower {S}ources},
  volume={566},
  pages={232861},
  doi={10.1016/j.jpowsour.2023.232861},
  year={2023}
}

@article{Schneider2019,
  title={Degradation phenomena of bismuth-modified felt electrodes in {VRFB} studied by electrochemical impedance spectroscopy},
  author={Schneider, Jonathan and Bulczak, Eduard and El-Nagar, Gumaa A and Gebhard, Marcus and Kubella, Paul and Schnucklake, Maike and Fetyan, Abdulmonem and Derr, Igor and Roth, Christina},
  journal={Batteries},
  volume={5},
  number={1},
  pages={16},
  doi={10.3390/batteries5010016},
  year={2019}
}

@article{Choi2017,
  title={A review of vanadium electrolytes for vanadium redox flow batteries},
  author={Choi, Chanyong and Kim, Soohyun and Kim, Riyul and Choi, Yunsuk and Kim, Soowhan and Jung, Ho-young and Yang, Jung Hoon and Kim, Hee-Tak},
  journal={Renew. {S}ustain. {E}nergy {R}ev.},
  volume={69},
  pages={263--274},
  doi={10.1016/j.rser.2016.11.188},
  year={2017}
}

@article{Tian2023,
  title={A review of electrolyte additives in vanadium redox flow batteries},
  author={Tian, Wenxin and Du, Hao and Wang, Jianzhang and Weigand, Jan J and Qi, Jian and Wang, Shaona and Li, Lanjie},
  journal={Materials},
  volume={16},
  number={13},
  pages={4582},
  doi={10.3390/ma16134582},
  year={2023}
}

@article{SkyllasKazacos2016,
  title={Vanadium electrolyte studies for the vanadium redox battery—{a} review},
  author={Skyllas-Kazacos, Maria and Cao, Liuyue and Kazacos, Michael and Kausar, Nadeem and Mousa, Asem},
  journal={Chem{S}us{C}hem},
  volume={9},
  number={13},
  pages={1521--1543},
  doi={10.1002/cssc.201600102},
  year={2016}
}

@article{Wu2014,
  title={Electrolytes for vanadium redox flow batteries},
  author={Wu, Xiongwei and Liu, Jun and Xiang, Xiaojuan and Zhang, Jie and Hu, Junping and Wu, Yuping},
  journal={Pure {A}ppl. {C}hem.},
  volume={86},
  number={5},
  pages={661--669},
  doi={10.1515/pac-2013-1213},
  year={2014}
}

@article{Fetyan2019,
  title={Detrimental role of hydrogen evolution and its temperature-dependent impact on the performance of vanadium redox flow batteries},
  author={Fetyan, Abdulmonem and El-Nagar, Gumaa A and Lauermann, Iver and Schnucklake, Maike and Schneider, Jonathan and Roth, Christina},
  journal={J. {E}nergy {C}hem.},
  volume={32},
  pages={57--62},
  doi={10.1016/j.jechem.2018.06.010},
  year={2019}
}

@article{Zhang2015,
  title={Effects of operating temperature on the performance of vanadium redox flow batteries},
  author={Zhang, Cheng and Zhao, TS and Xu, Qian and An, Liang and Zhao, Gang},
  journal={Appl. {E}nergy},
  volume={155},
  pages={349--353},
  doi={10.1016/j.apenergy.2015.06.002},
  year={2015}
}

@article{Wei2017,
  title={In-situ investigation of hydrogen evolution behavior in vanadium redox flow batteries},
  author={Wei, Lei and Zhao, TS and Xu, Qian and Zhou, XL and Zhang, ZH},
  journal={Appl. {E}nergy},
  volume={190},
  pages={1112--1118},
  doi={10.1016/j.apenergy.2017.01.039},
  year={2017}
}

@article{Ye2024,
  title={Gas evolution induced vicious cycle between bubble trapping and flow choking in redox flow battery stacks},
  author={Ye, Qiang and Dai, Jincheng and Cheng, Ping and Zhao, Tianshou},
  journal={Int. {J}. {H}eat {M}ass {T}ransf.},
  volume={221},
  pages={125100},
  doi={10.1016/j.ijheatmasstransfer.2023.125100},
  year={2024}
}

@article{Eifert2020,
  title={Synchrotron {X}-ray radiography and tomography of vanadium redox flow batteries—cell design, electrolyte flow geometry, and gas bubble formation},
  author={Eifert, L{\'a}szl{\'o} and Bevilacqua, Nico and K{\"o}ble, Kerstin and Fahy, Kieran and Xiao, Liusheng and Li, Min and Duan, Kangjun and Bazylak, Aimy and Sui, Pang-Chieh and Zeis, Roswitha},
  journal={Chem{S}us{C}hem},
  volume={13},
  number={12},
  pages={3154--3465},
  doi={10.1002/cssc.202000541},
  year={2020}
}

@article{Koble2023,
  title={Revealing the Multifaceted Impacts of Electrode Modifications for Vanadium Redox Flow Battery Electrodes},
  author={K{\"o}ble, Kerstin and Schilling, Monja and Eifert, L{\'a}szl{\'o} and Bevilacqua, Nico and Fahy, Kieran F and Atanassov, Plamen and Bazylak, Aimy and Zeis, Roswitha},
  journal={{ACS} {A}ppl. {M}ater. {I}nterfaces},
  volume={15},
  number={40},
  pages={46775--46789},
  doi={10.1021/acsami.3c07940},
  year={2023}
}

@article{Koble2024,
  title={Insights into the hydrogen evolution reaction in vanadium redox flow batteries: {A} synchrotron radiation based {X}-ray imaging study},
  author={K{\"o}ble, Kerstin and Ershov, Alexey and Duan, Kangjun and Schilling, Monja and Rampf, Alexander and Cecilia, Angelica and Farag{\'o}, Tom{\'a}{\v{s}} and Zuber, Marcus and Baumbach, Tilo and Zeis, Roswitha},
  journal={J. {E}nergy {C}hem.},
  volume={91},
  pages={132--144},
  doi={10.1016/j.jechem.2023.12.010},
  year={2024}
}

@article{Chen2017,
  title={Pore-scale study of multiphase reactive transport in fibrous electrodes of vanadium redox flow batteries},
  author={Chen, Li and He, YaLing and Tao, Wen-Quan and Zelenay, Piotr and Mukundan, Rangachary and Kang, Qinjun},
  journal={Electrochim. {A}cta},
  volume={248},
  pages={425--439},
  doi={10.1016/j.electacta.2017.07.086},
  year={2017}
}

@article{Zhang2018,
  title={The effect of wetting area in carbon paper electrode on the performance of vanadium redox flow batteries: A three-dimensional lattice {Boltzmann} study},
  author={Zhang, Duo and Cai, Qiong and Taiwo, Oluwadamilola O and Yufit, Vladimir and Brandon, Nigel P and Gu, Sai},
  journal={Electrochim. {A}cta},
  volume={283},
  pages={1806--1819},
  doi={10.1016/j.electacta.2018.07.027},
  year={2018}
}

@article{Li2020b,
  title={Mesoscopic modeling and characterization of the porous electrodes for vanadium redox flow batteries},
  author={Li, Min and Bevilacqua, Nico and Zhu, Lijun and Leng, Wengliang and Duan, Kangjun and Xiao, Liusheng and Zeis, Roswitha and Sui, Pang-Chieh},
  journal={J. {E}nergy {S}torage},
  volume={32},
  pages={101782},
  doi={10.1016/j.est.2020.101782},
  year={2020}
}

@article{Leclaire2017b,
  title={Generalized three-dimensional lattice {Boltzmann} color-gradient method for immiscible two-phase pore-scale imbibition and drainage in porous media},
  author={Leclaire, S{\'e}bastien and Parmigiani, Andrea and Malaspinas, Orestis and Chopard, Bastien and Latt, Jonas},
  journal={Phys. {R}ev. {E}},
  volume={95},
  number={3},
  pages={033306},
  doi={10.1103/PhysRevE.95.033306},
  year={2017}
}

@article{scheel2024,
  title={Enhancement of bubble transport in porous electrodes and catalysts},
  author={Scheel, Thomas and Malgaretti, Paolo and Harting, Jens},
  journal={J. {C}hem. {P}hys.},
  volume={160},
  number={19},
  year={2024},
  doi={10.1063/5.0206381},
  pages={194706}
}

@article{Schweiss2016,
  title={Parasitic hydrogen evolution at different carbon fiber electrodes in vanadium redox flow batteries},
  author={Schweiss, Ruediger and Pritzl, Alexander and Meiser, Christian},
  journal={J. {E}lectrochem. {S}oc.},
  volume={163},
  number={9},
  pages={A2089},
  doi={10.1149/2.1281609jes},
  year={2016}
}

@article{shan1993lattice,
  title={Lattice {B}oltzmann model for simulating flows with multiple phases and components},
  author={Shan, Xiaowen and Chen, Hudong},
  journal={Phys. {R}ev. {E}},
  volume={47},
  number={3},
  pages={1815},
  doi={10.1103/PhysRevE.47.1815},
  year={1993}
}

@article{scheel2023viscous,
  title={Viscous to inertial coalescence of liquid lenses: {A} lattice {B}oltzmann investigation},
  author={Scheel, Thomas and Xie, Qingguang and Sega, Marcello and Harting, Jens},
  journal={Phys. {R}ev. {F}luids},
  volume={8},
  number={7},
  pages={074201},
  doi={10.1103/PhysRevFluids.8.074201},
  year={2023}
}

@article{guo2002discrete,
  title={Discrete lattice effects on the forcing term in the lattice {B}oltzmann method},
  author={Guo, Zhaoli and Zheng, Chuguang and Shi, Baochang},
  journal={Phys. {R}ev. {E}},
  volume={65},
  number={4},
  pages={046308},
  doi={10.1103/PhysRevE.65.046308},
  year={2002}
}

@article{gunstensen1991lattice,
  title={Lattice {B}oltzmann model of immiscible fluids},
  author={Gunstensen, Andrew K and Rothman, Daniel H and Zaleski, St{\'e}phane and Zanetti, Gianluigi},
  journal={Phys. {R}ev. {A}},
  volume={43},
  number={8},
  pages={4320},
  doi={10.1103/PhysRevA.43.4320},
  year={1991}
}

@article{narvaez2013creeping,
  title={From creeping to inertial flow in porous media: a lattice {B}oltzmann--finite element study},
  author={Narv{\'a}ez, Ariel and Yazdchi, Kazem and Luding, Stefan and Harting, Jens},
  journal={J. {S}tat. {M}ech. {T}heory {E}xp.},
  volume={2013},
  number={02},
  pages={P02038},
  doi={10.1088/1742-5468/2013/02/P02038},
  year={2013}
}

@article{latva2005static,
  title={Static contact angle in lattice {B}oltzmann models of immiscible fluids},
  author={Latva-Kokko, M and Rothman, Daniel H},
  journal={Phys. {R}ev. {E}},
  volume={72},
  number={4},
  pages={046701},
  doi={10.1103/PhysRevE.72.046701},
  year={2005}
}

@article{gur2018review,
  title={Review of electrical energy storage technologies, materials and systems: challenges and prospects for large-scale grid storage},
  author={G{\"u}r, Turgut M},
  journal={Energy {E}nviron. {S}ci.},
  volume={11},
  number={10},
  pages={2696--2767},
  doi={10.1039/C8EE01419A},
  year={2018}
}

@article{zhang2019progress,
  title={Progress and perspectives of flow battery technologies},
  author={Zhang, Huamin and Lu, Wenjing and Li, Xianfeng},
  journal={Electrochem. {E}nergy {R}ev.},
  volume={2},
  pages={492--506},
  doi={10.1007/s41918-019-00047-1},
  year={2019}
}

@article{dunn2011electrical,
  title={Electrical energy storage for the grid: a battery of choices},
  author={Dunn, Bruce and Kamath, Haresh and Tarascon, Jean-Marie},
  journal={Science},
  volume={334},
  number={6058},
  pages={928--935},
  doi={10.1126/science.1212741},
  year={2011}
}

@article{skyllas2011progress,
  title={Progress in flow battery research and development},
  author={Skyllas-Kazacos, Maria and Chakrabarti, MH and Hajimolana, SA and Mjalli, FS and Saleem, M},
  journal={J. {E}lectrochem. {S}oc.},
  volume={158},
  number={8},
  pages={R55},
  doi={10.1149/1.3599565},
  year={2011}
}

@article{minke2018materials,
  title={Materials, system designs and modelling approaches in techno-economic assessment of all-vanadium redox flow batteries--{A} review},
  author={Minke, Christine and Turek, Thomas},
  journal={J. {P}ower {S}ources},
  volume={376},
  pages={66--81},
  doi={10.1016/j.jpowsour.2017.11.058},
  year={2018}
}

@article{shi2019recent,
  title={Recent development of membrane for vanadium redox flow battery applications: {A} review},
  author={Shi, Yu and Eze, Chika and Xiong, Binyu and He, Weidong and Zhang, Han and Lim, Tuti Mariana and Ukil, A and Zhao, Jiyun},
  journal={Appl. {E}nergy},
  volume={238},
  pages={202--224},
  doi={10.1016/j.apenergy.2018.12.087},
  year={2019}
}

@article{wu2018electrocatalysis,
  title={Electrocatalysis at electrodes for vanadium redox flow batteries},
  author={Wu, Yuping and Holze, Rudolf},
  journal={Batteries},
  volume={4},
  number={3},
  pages={47},
  doi={10.3390/batteries4030047},
  year={2018}

}

@article{wang2018reduction,
  title={The reduction reaction kinetics of vanadium ({V}) in acidic solutions on a platinum electrode with unusual difference compared to carbon electrodes},
  author={Wang, Wenjun and Fan, Xinzhuang and Qin, Ye and Liu, Jianguo and Yan, Chuanwei and Zeng, Chaoliu},
  journal={Electrochim. {A}cta},
  volume={283},
  pages={1313--1322},
  doi={10.1016/j.electacta.2018.07.071},
  year={2018},

}

@article{li2022techno,
  title={Techno-economic analysis of non-aqueous hybrid redox flow batteries},
  author={Li, Zhiguang and Fang, Xiaoting and Cheng, Lei and Wei, Xiaoliang and Zhang, Lu},
  journal={J. {P}ower {S}ources},
  volume={536},
  pages={231493},
  doi={10.1016/j.jpowsour.2022.231493},
  year={2022}

}

@article{baritto2022development,
  title={The development of a techno-economic model for the assessment of vanadium recovery from bitumen upgrading spent catalyst},
  author={Baritto, M and Oni, AO and Kumar, A},
  journal={J. {C}lean. {P}rod.},
  volume={363},
  pages={132376},
  doi={10.1016/j.jclepro.2022.132376},
  year={2022}

}

@article{rodby2020assessing,
  title={Assessing the levelized cost of vanadium redox flow batteries with capacity fade and rebalancing},
  author={Rodby, Kara E and Carney, Thomas J and Gandomi, Yasser Ashraf and Barton, John L and Darling, Robert M and Brushett, Fikile R},
  journal={J. {P}ower {S}ources},
  volume={460},
  pages={227958},
  doi={10.1016/j.jpowsour.2020.227958},
  year={2020}

}

@article{sheng2018robust,
  title={Robust electrodes with maximized spatial catalysis for vanadium redox flow batteries},
  author={Sheng, Hang and Ma, Qiang and Yu, Jin-Gang and Zhang, Xu-Dong and Zhang, Wei and Yin, Ya-Xia and Wu, Xiongwei and Zeng, Xian-Xiang and Guo, Yu-Guo},
  journal={{ACS} {A}ppl. {M}ater. {I}nterfaces},
  volume={10},
  number={45},
  pages={38922--38927},
  year={2018},
  doi={10.1021/acsami.8b13778},
  publisher={ACS Publications}
}

@article{xiao2020pore,
  title={Pore-scale characterization and simulation of porous electrode material for vanadium redox flow battery: effects of compression on transport properties},
  author={Xiao, Liusheng and Luo, Maji and Zhu, Lijun and Duan, Kangjun and Bevilacqua, Nico and Eifert, L{\'a}szl{\'o} and Zeis, Roswitha and Sui, Pang-Chieh},
  journal={J. {E}lectrochem. {S}oc.},
  volume={167},
  number={11},
  pages={110545},
  year={2020},
  doi={10.1149/1945-7111/aba4e3},
  publisher={IOP Publishing}
}

@article{jeong2023enhancing,
  title={Enhancing Vanadium redox flow batteries performance through local compression ratio adjustment using stiffness gradient carbon felt electrodes},
  author={Jeong, Kwang Il and Lim, Su Hyun and Hong, Hyunsoo and Jeong, Jae-Moon and Kim, Won Vin and Kim, Seong Su},
  journal={Appl. {M}ater. {T}oday},
  volume={35},
  pages={101928},
  year={2023},
  doi={10.1016/j.apmt.2023.101928},
  publisher={Elsevier}
}

@article{hsieh2019effect,
  title={Effect of compression ratio of graphite felts on the performance of an all-vanadium redox flow battery},
  author={Hsieh, Chin-Lung and Tsai, Po-Hong and Hsu, Ning-Yih and Chen, Yong-Song},
  journal={Energies},
  volume={12},
  number={2},
  pages={313},
  year={2019},
  doi={10.3390/en12020313},
  publisher={MDPI}
}

@article{bromberger2014model,
  title={A model for all-vanadium redox flow batteries: introducing electrode-compression effects on voltage losses and hydraulics},
  author={Bromberger, Kolja and Kaunert, Johannes and Smolinka, Tom},
  journal={Energy {T}echnol.},
  volume={2},
  number={1},
  pages={64--76},
  year={2014},
  doi={10.1002/ente.201300114},
  publisher={Wiley Online Library}
}

@article{wang2023numerical,
  title={Numerical study of vanadium redox flow battery with gradient porosity induced by electrode compression},
  author={Wang, Q and Yan, RJ and Chen, JC and Jiang, ZY and Qu, ZG},
  journal={J. {E}nergy {S}torage},
  volume={72},
  pages={108465},
  year={2023},
  doi={10.1016/j.est.2023.108465},
  publisher={Elsevier}
}

@article{qin2022measurement,
  title={Measurement and accurate prediction of surface tension for {VOSO$_4$}-{H$_2$SO$_4$}-{H$_2$O} ternary electrolyte system at high-concentration in vanadium redox flow batteries},
  author={Qin, Ye and Qi, Peixia and Zhao, Jinling and Li, Xiangrong and Wang, Na and Li, Qingpeng and Ge, Jing and Liu, Jianguo and Yang, Jiazhen and Yan, Chuanwei},
  journal={J. {M}ol. {L}iq.},
  volume={365},
  pages={120079},
  year={2022},
  doi = {10.1016/j.molliq.2022.120079},
  publisher={Elsevier}
}

@article{chen2019inertial,
  title={Inertial effects during the process of supercritical {CO$_2$} displacing brine in a sandstone: {L}attice {B}oltzmann simulations based on the continuum-surface-force and geometrical wetting models},
  author={Chen, Yu and Valocchi, Albert J and Kang, Qinjun and Viswanathan, Hari S},
  journal={Water {R}esour. {R}es.},
  volume={55},
  number={12},
  pages={11144--11165},
  year={2019},
  doi = {10.1029/2019WR025746},
  publisher={Wiley Online Library}
}

@article{colliard2025advancing,
  title={Advancing vanadium redox flow battery analysis: a deep learning approach for high-throughput 3{D} visualization and bubble quantification},
  author={Colliard-Granero, Andr{\'e} and Duan, Kangjun and Zeis, Roswitha and Eikerling, Michael and Malek, Kourosh and Eslamibidgoli, Mohammad J},
  journal={Digital {D}iscov.},
  year={2025},
  doi={10.1039/D5DD00158G},
  publisher={Royal Society of Chemistry}
}

@article{ma2024evaluation,
  title={Evaluation of the effect of hydrogen evolution reaction on the performance of all-vanadium redox flow batteries},
  author={Ma, Tao and Huang, Zebo and Xie, Xing and Li, Bin},
  journal={Electrochim. {A}cta},
  volume={504},
  pages={144895},
  year={2024},
  doi = {10.1016/j.electacta.2024.144895},
  publisher={Elsevier}
}

@article{chai2024double,
  title={A double-spiral flow channel of vanadium redox flow batteries for enhancing mass transfer and reducing pressure drop},
  author={Chai, Yuwei and Qu, Dawei and Fan, Luyan and Zheng, Yating and Yang, Fan},
  journal={J. {E}nergy {S}torage},
  volume={78},
  pages={110278},
  year={2024},
  doi = {10.1016/j.est.2023.110278},
  publisher={Elsevier}
}

@article{schilling2024investigating,
  title={Investigating the influence of treatments on carbon felts for vanadium redox flow batteries},
  author={Schilling, Monja and Eifert, L{\'a}szl{\'o} and K{\"o}ble, Kerstin and Jaugstetter, Maximilian and Bevilacqua, Nico and Fahy, Kieran F and Tschulik, Kristina and Bazylak, Aimy and Zeis, Roswitha},
  journal={Chem{S}us{C}hem},
  volume={17},
  number={1},
  pages={e202301063},
  year={2024},
  doi = {10.1002/cssc.202301063},
  publisher={Wiley Online Library}
}

@article{ye2024vanadium,
  title={Vanadium redox flow battery: review and perspective of 3{D} electrodes},
  author={Ye, Lingzhi and Qi, Shaotian and Cheng, Tukang and Jiang, Yingqiao and Feng, Zemin and Wang, Mingyong and Liu, Yongguang and Dai, Lei and Wang, Ling and He, Zhangxing},
  journal={{ACS} {N}ano},
  volume={18},
  number={29},
  pages={18852--18869},
  year={2024},
  doi = {10.1021/acsnano.4c06675},
  publisher={ACS Publications}
}

@article{nath2025reaction,
  title={Reaction-limited evaporation for the color-gradient lattice {B}oltzmann model},
  author={Nath, Gaurav and Aouane, Othmane and Harting, Jens},
  journal={J. {C}hem. {P}hys.},
  volume={162},
  number={11},
  year={2025},
  doi = {10.1063/5.0253799},
  publisher={AIP Publishing}
}

@article{sedahmed2024wetting,
  title={Wetting and pressure gradient performance in a lattice {B}oltzmann color gradient model},
  author={Sedahmed, Mahmoud and Coelho, Rodrigo CV},
  journal={Phys. {F}luids},
  volume={36},
  number={9},
  pages={092117},
  year={2024},
  doi = {10.1063/5.0228835},
  publisher={AIP Publishing LLC}
}

@article{zahid2025review,
  title={Review of the color gradient lattice {B}oltzmann method for simulating multi-phase flow in porous media: {V}iscosity, gradient calculation, and fluid acceleration},
  author={Zahid, Fizza and Cunningham, Jeffrey A},
  journal={Fluids},
  volume={10},
  number={5},
  pages={128},
  year={2025},
  doi = {10.3390/fluids10050128},
  publisher={MDPI}
}

\end{document}